%% file: sophie11_vArXiv.tex
\documentclass[traditabstract]{aa}

\usepackage{graphicx}
\usepackage{tikz}

\usepackage{txfonts}
\usepackage{natbib}
\usepackage{epsf}
\usepackage{rotating}
\usepackage{graphics}
\usepackage{verbatim}
\usepackage{longtable}

\newcommand{\logR}{$\log{(R'_{\rm HK})}$}

\def \MJ{M$_{\mathrm{Jup}}$}

\def \kms{km\,s$^{-1}$}
\def \ms{m\,s$^{-1}$}
\def \1s{$1\,\sigma$}

\def \t0{T$_0$}

\newcommand{\teta}{\boldsymbol{\theta}}

\begin{document}

\title{The SOPHIE search for northern extrasolar planets\thanks{Based on observations collected with the {\it SOPHIE}  spectrograph on the 1.93-m telescope at Observatoire de Haute-Provence (CNRS), France by the {\it SOPHIE}  Consortium  (programme 07A.PNP.CONS to 15A.PNP.CONS).}}

\subtitle{XI. Three new companions and an orbit update: Giant planets in the habitable zone.}

\author{
R.~F.~D\'iaz\inst{1} \and 
J.~Rey\inst{1} \and 
O.~Demangeon\inst{2}\and
G.~H\'ebrard\inst{3,4} \and 
I.~Boisse\inst{2} \and
L.~Arnold\inst{4}\and
N.~Astudillo-Defru\inst{1}\and
J.-L.~Beuzit\inst{5,6}\and
X.~Bonfils\inst{5,6}\and 
S.~Borgniet\inst{5,6}\and
F.~Bouchy\inst{1,2}\and
V.~Bourrier\inst{1}\and
B.~Courcol\inst{2}\and
M.~Deleuil\inst{2}\and
X.~Delfosse\inst{5,6}\and
D.~Ehrenreich\inst{1}\and
T.~Forveille\inst{5,6}\and 
A.-M.~Lagrange\inst{5,6}\and
M.~Mayor\inst{1}\and
C.~Moutou\inst{2,7}\and 
F.~Pepe\inst{1}\and
D.~Queloz\inst{1,8}\and
A.~Santerne\inst{9} \and
N.~C.~Santos\inst{9,10}\and
J.~Sahlmann\inst{11}\and 
D.~S\'egransan\inst{1}\and 
S.~Udry\inst{1}\and 
P.~A.~Wilson\inst{3}
}

 \offprints{R.F D\'iaz (rodrigo.diaz@unige.ch)}

\institute{
Observatoire Astronomique de l'Universit\'e de Gen\`eve,  51 Chemin des Maillettes, 1290 Versoix, Switzerland\and
Aix Marseille Universit\'e, CNRS, LAM (Laboratoire d'Astrophysique de Marseille) UMR 7326, 13388 Marseille, France \and
Institut d'Astrophysique de Paris, UMR7095 CNRS, Universit\'e Pierre \& Marie Curie, 98bis boulevard Arago, 75014 Paris, France \and 
Observatoire de Haute Provence, CNRS, Aix Marseille Universit\'e, Institut Pyth\'eas UMS 3470, 04870 Saint-Michel-l'Observatoire, France\and 
Univ. Grenoble Alpes, IPAG, F-38000 Grenoble, France \and
CNRS, IPAG, F-38000 Grenoble, France \and
Canada-France-Hawaii Telescope Corporation, 65-1238 Mamalahoa Hwy, Kamuela, HI 96743, USA\and
Cavendish Laboratory, J J Thomson Avenue, Cambridge CB3 0HE, UK\and 
Instituto de Astrof\'isica e Ci\^encias do Espa\c{c}o, Universidade do Porto, CAUP, Rua das Estrelas, 4150-762 Porto, Portugal \and
Departamento de F\'isica e Astronomia, Faculdade de Ci\^encias, Universidade do Porto, Rua do Campo Alegre, 4169-007 Porto, Portugal \and
European Space Agency, European Space Astronomy Centre, P.O. Box 78, Villanueva de la Canada, 28691 Madrid, Spain
}

 \date{Received TBC; accepted TBC}
      
\abstract{We report the discovery of three new substellar companions to solar-type stars, HD191806, HD214823, and HD221585, based on radial velocity measurements obtained at the Haute-Provence Observatory. Data from the SOPHIE spectrograph are combined with observations acquired with its predecessor, ELODIE, to detect and characterise the orbital parameters of three new gaseous giant and brown dwarf candidates. Additionally, we combine SOPHIE data with velocities obtained at the Lick Observatory to improve the parameters of an already known giant planet companion, HD16175 b. Thanks to the use of different instruments, the data sets of all four targets span more than ten years. Zero-point offsets between instruments are dealt with using Bayesian priors to incorporate the information we possess on the SOPHIE/ELODIE offset based on previous studies.

The reported companions have orbital periods between three and five years and minimum masses between 1.6 \MJ\ and 19 \MJ. Additionally, we find that the star HD191806 is experiencing a secular acceleration of over 11 \ms\ per year, potentially due to an additional stellar or substellar companion. A search for the astrometric signature of these companions was carried out using Hipparcos data. No orbit was detected, but a significant upper limit to the companion mass can be set for HD221585, whose companion must be substellar.

With the exception of HD191806 b, the companions are located within the habitable zone of their host star. Therefore, satellites orbiting these objects could be a propitious place for life to develop.
}

\authorrunning{D\'iaz et al.}
\titlerunning{Long-period Jupiter-mass planets.}

\keywords{planetary systems -- techniques: radial velocities -- stars: brown dwarfs -- stars: individual: \object{HD16175}, \object{HD191806}, \object{HD214823}, \object{HD221585}}

\maketitle

\section{Introduction \label{sect.intro}}

As the baselines of radial velocity surveys move from years to decades, it becomes possible to probe regions progressively farther away from the targeted stars. 
With over 20 years of radial velocity observations under the belt, this has in fact already been achieved for giant planets. As of December 2015, the exoplanets.org database \citep{han2014} listed over a hundred planets detected with semi-major axes larger than 2 au. Most of these orbit solar-type stars and,  with two notable exceptions, namely HD10180 h \citep{lovis2011} and HD204941 b \citep{dumusque2011c}, all of these planets have masses larger than Saturn.


The discovery of giant planetary companions on long-period orbits is important for a number of reasons: Firstly,  probing the architecture of extrasolar systems out to the ice line and beyond provides strong constraints on planet formation and evolution. Secondly, it allows us to put our own solar system in context. Thirdly, these giant planets are likely to play an important role in the short-term dynamics of the systems. For example, an unseen massive companion in an outer orbit can bias the mass determinations obtained via transit timing variations of small-mass planets in close-in compact systems, the likes of which have been abundantly discovered by the \emph{Kepler} mission \citep[e.g.][]{rowe2014}. Finally, on account of the role of Jupiter in the impact rate on Earth of asteroids and comets, and the consequent effect on habitability  \citep[see ][and references therein]{laakso2006, hornerjones2008, hornerjones2010}. Additionally, although gaseous planets are most likely incapable of harbouring life, they could be orbited by rocky satellites that are analogous to the Galilean satellites of Jupiter or larger. If the giant planet orbits in the habitable zone of its star, these moons would constitute a suitable place for life to develop \citep{williams1997, portergrundy2011, heller2014, hellerpudritz2015}. Therefore, establishing whether a gaseous planet is in the habitable zone \citep{kasting1993, kopparapu2013} bears importance for future astrobiological studies.

Two factors act as the main limitation to the endeavour of unveiling such planets. Firstly, no single unmodified RV instrument has observed uninterrupted for 20 years; ELODIE \citep{baranne1996} was replaced with SOPHIE in 2006 \citep{perruchot2008, bouchy2009}. In turn SOPHIE was upgraded in 2011 \citep{bouchy2013, perruchot2011}. The HIRES instrument was upgraded in 2004 \citep{vogt1994}, and CORALIE was upgraded in 2007 and  2014 \citep{queloz2000, segransan2010}. The planet search programme at the 2.4-m telescope of the McDonald Observatory went through three instrumental phases, the latter of which started in 1998 \citep{hatzes2000, endl2004}. Even HARPS, which remained untouched for over ten years, has recently been upgraded \citep{mayor2003, locurto2015}. This implies that instrument offsets have to be treated cautiously to avoid introducing spurious signals that pass for long-period giant planets. This is particularly important when no overlap between the instruments exist, as is the case for most of the examples cited above. 

Secondly, many solar-type stars, even the most inactives, exhibit activity cycles or variations with typical timescales between a few years and a couple of decades \citep{baliunas95, lovis2011b}. Long-term activity variations are related to global changes in the convective pattern of the star \citep{lindegrendravins2003, meunier2010, meunierlagrange2013}, and produce radial velocity signals clearly correlated with activity proxies such as \logR\ \citep{noyes84} and the width and asymmetry of the mean spectral line \citep{dravins1982, santos2010, lovis2011b, dumusque2011c, diaz2016a}. These  signals can be mistaken with long-period planets, and if not fully corrected, can contaminate the radial velocity time series at higher frequencies. 

In this article, we tackle these issues by modelling radial velocity time series using a Bayesian approach. Information such as the calibrations of instrument offsets are incorporated into the model through the use of Bayesian priors. We report the detection of three new long-period giant companions around solar-type stars based on data obtained with the ELODIE and SOPHIE spectrographs. One of these companions has a minimum mass of around 19 \MJ and may in fact be a brown dwarf. Additionally, we improved upon the orbital and physical parameters of a known giant planet candidate by combining the original discovery data with new measurements from ELODIE and SOPHIE. The paper is organised as follows: In Sect.~\ref{sect.observations} we describe the observations and the techniques used to reduce the data. Section~\ref{sect.stellarparams} describes the stellar hosts, their basic physical properties and their activity level over the time span of our observations. We discuss the presence of activity cycles here. Section~\ref{sect.bisector} presents the results of the analysis of the bisector of the stellar mean line, and in Sect.~\ref{sect.analysis} we describe our model and the technique to sample the posterior distribution of the model parameters. The results are presented individually for each target in Sect.~\ref{sect.results}. We study the position of the candidates with respect to the habitable zone and the habitability of potential satellites in Sect.~\ref{sect.hz} and, finally, we give a summary of the results in Sect.~\ref{sect.discussion}.

\section{Observations and data reduction \label{sect.observations}}

\begin{table*}[t]
\center
\caption{Basic characteristics of the ELODIE, SOPHIE, and SOPHIE+ observations of the four target stars. $N$ is the total number of spectra. The average uncertainty due to photon noise and calibration error, $<\sigma_\mathrm{RV}>$, is also provided. \label{table.obslog}}
\begin{tabular}{l l cc  ccc  cc }
\hline
\hline
Target & instrument  & $N$ & \multicolumn{2}{c}{Dates} & time span & $<\sigma_\mathrm{RV}>$\\
 & & & start & end & [yr] &  [m/s]\\
\hline
HD16175    & ELODIE & 3 & 2004-10-03 & 2004-11-24    & 0.14 & 9.0\\
                  & SOPHIE & --   & --            & --                  & --    &  -- \\
                  & SOPHIE+ & 25 & 2011-09-02 & 2015-09-03 & 4.0 & 4.5\\
\hline
HD191806  & ELODIE &6 & 2004-09-26 & 2006-08-12 & 1.9 & 12.9\\
                   & SOPHIE &28 & 2006-12-23 & 2011-04-08 & 4.3 & 4.0\\
                   & SOPHIE+&18 & 2011-09-02 & 2015-05-11 & 3.7 & 4.7\\
\hline
HD214823  & ELODIE& 5  &  2005-08-04 & 2006-08-13 &  1.0 & 21.2\\
                  & SOPHIE& 13 & 2008-07-20 & 2011-05-21 & 2.8 & 6.0 \\
                  & SOPHIE+& 11& 2011-08-03 & 2015-06-30 & 3.9 & 5.8\\
\hline
HD221585  & ELODIE & 10 & 2004-09-28 & 2006-07-16 & 1.8 & 9.0\\
                   & SOPHIE & 5 & 2008-08-16 & 2010-08-21 & 2.0 & 3.1\\
                   & SOPHIE+ & 20 & 2011-08-17 & 2015-08-24 & 4.0 & 3.4\\ 
\hline
\end{tabular}
\end{table*}

The targets were observed as part of the volume-limited SOPHIE survey for giant planets (sub-programme 2, or SP2; \citealt{bouchy2009}). SOPHIE is a fibre-fed, cross-dispersed \emph{echelle} spectrograph mounted at the 1.93-m telescope of the Haute-Provence Observatory. Its dispersive elements are kept at constant pressure and it is installed in a temperature stabilised environment \citep{perruchot2008} to provide high-precision radial velocity measurements over long timescales. Observations of SP2 targets are carried out in the high-resolution mode of the instrument, which provides resolving power R = 75'000. In addition to the target fibre, a second fibre monitors sky brightness variations, particularly because of the scattered moonlight.

Sub-programme 2 targets 2300 dwarf stars at distance < 60 pc and with B-V between 0.35 and 1.0. Most of these stars (about 1900) have already been observed with SOPHIE \citep{hebrard2016}. The programme employs medium-precision measurements (around 3 m/s) that permitted the detection of around 50 substellar companions, including giant planets in multi-planetary systems \citep[e.g.~][]{hebrard2010b, moutou2014} and brown dwarf companions \citep{diaz2012, wilson2016}.  The observing strategy consists in acquiring spectra with a constant signal-to-noise ratio (S/N) of around 50 to minimise the effects of the charge transfer inefficiency \citep{bouchy2009c}. This is usually achieved in a few minutes of integration time, but as the SP2 is also used as a backup programme for the SOPHIE programmes that require more precise measurements, some exposures during poor weather conditions are much longer. 

The spectra were reduced using the SOPHIE pipeline described by \citet{bouchy2009}. The radial velocity (RV) was measured by weighted cross correlation of the extracted two-dimensional echelle spectra with a spectral mask corresponding to the spectral type of the observed star \citep{baranne1996, pepe2002}. A Gaussian curve was fitted to the cross-correlation function (CCF) averaged over all spectral orders to produce the RV measurement. Additionally, we obtained the bisector velocity span of the CCF as described by \citet{queloz2001}. The measurements are listed in Tables~\ref{table.rvHD16175} through \ref{table.rvHD221585}\footnote{These tables contain the raw measurements. The offset between the instrument zero-points was not included (see Sect.~\ref{sect.model}).}

One of the main systematic effects in SOPHIE data was due to insufficient scrambling of the fibre link and the high sensitivity of the spectrograph to illumination variations. As a consequence, seeing changes at the fibre input produced changes in the measured radial velocity. This so-called seeing effect is described in \citet{boisse2010, boisse2011b}, \citet{diaz2012}, and \citet{bouchy2013}. The seeing effect can be partially corrected for by measuring the RV on each half of spectral orders and using the difference of the obtained velocities to decorrelate the velocity measured with the entire detector \citep[see][]{bouchy2013}. This correction was applied to all data obtained with SOPHIE before June 2011. The seeing effect was dramatically reduced in June 2011 (BJD=2'455'730) with the introduction of octagonal-section fibres in the fibre link. The new fibre may have caused a zero-point shift, which is not yet fully characterised. In this article, we treat SOPHIE data obtained after June 2011 as if it were from a different instrument, which we refer to as SOPHIE+.  In spite of an important improvement in the SOPHIE RV precision, some systematic zero-point offsets are still observed. They are probably related to an incomplete thermal isolation and are corrected for as described by \citet{courcol2015}.

All of the targets were already observed with the ELODIE spectrograph \citep{baranne1996}, which is the predecessor of SOPHIE. However, they were not included in SOPHIE sub-programme 5, which follows up velocity trends and incomplete orbits identified with ELODIE \citep{bouchy2009, boisse2012b, bouchy2016}. The ELODIE RVs were also obtained by cross correlation with a numerical spectral mask. A velocity offset between ELODIE and SOPHIE exists and has been calibrated by \citet{boisse2012b} as a function of colour index $B-V$ using a sample of around 200 stars. These authors mentioned that a dependence on stellar metallicity must exist, but they claimed the effect is small and decided to neglect it.

Table~\ref{table.obslog} lists the main characteristics of the ELODIE, SOPHIE, and SOPHIE+ data sets. The data sets are plotted in Fig.~\ref{fig.rvcurves}. In all cases, clear RV variations exist, which are due to substellar orbiting companions, as we show below. The measured orbital and physical parameters are listed in Table~\ref{table.params}.

\begin{figure}[t]
\input{figures/HD16175_k1_sophiepc_lick_sophienoisemodel1_RVtime.pgf}
\input{figures/HD191806_k1d1_sophiepc_sophienoisemodel1_RVtime.pgf}
\input{figures/HD214823_k1_sophiepc_sophienoisemodel1_RVtime.pgf}
\input{figures/HD221585_k1_sophiepc_sophienoisemodel1_RVtime.pgf}
\caption{Radial velocity data and model fits. The green points represent ELODIE data, and the red empty and filled circles represent SOPHIE and SOPHIE+ data, respectively. The empty black points are Lick Observatory data. In the upper panel of each plot, the solid thick black curve is the maximum a posteriori physical model. The blue curves are the 95-\% highest density interval (HDI) for the complete model, computed as described in \citet{diaz2016a} and \citet{gregory2011}.  The model posterior mean is represented by solid thin
grey curve. The stellar drift or systemic velocity is shown as a thick grey line with the 95-\% (HDI) shown with dotted grey lines. \label{fig.rvcurves}}
\end{figure}

\section{Stellar parameters \label{sect.stellarparams}}

\input{StellarParams_sophie11}

\begin{figure*}[t]
\input{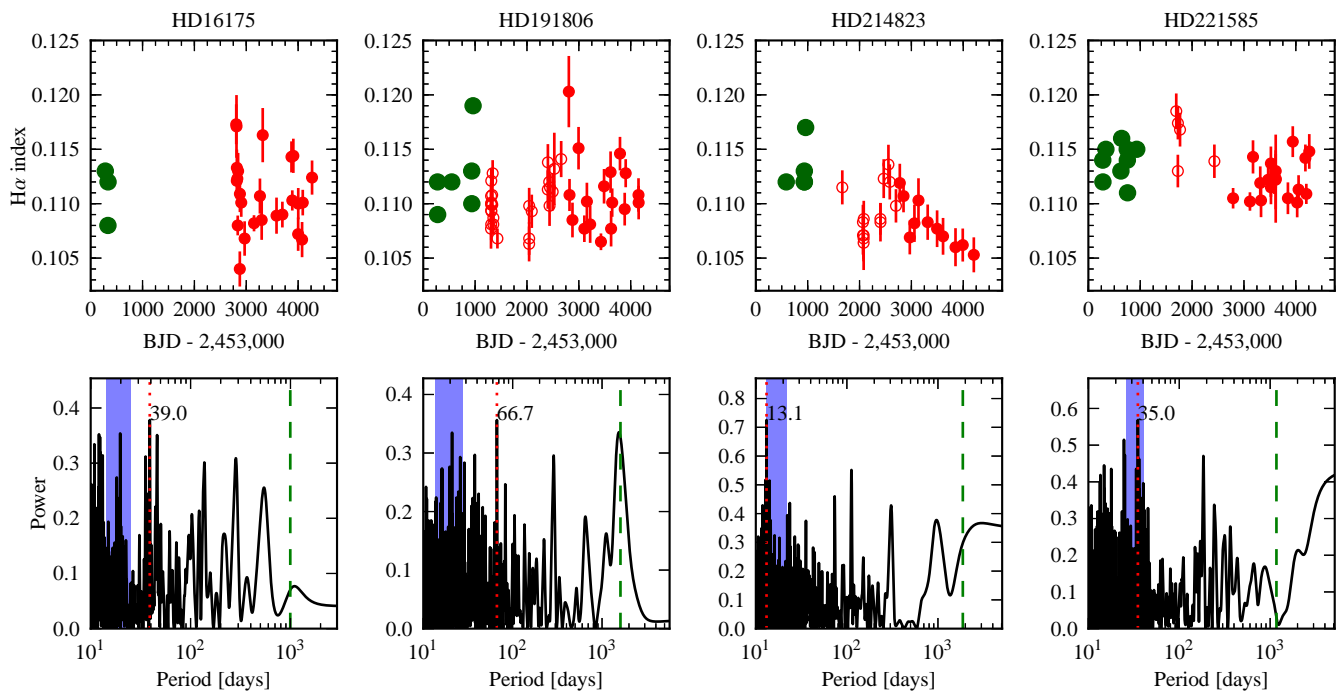}
\caption{\emph{Top row:} time series of the H$\alpha$ activity index. The green points correspond to ELODIE observations, and red empty symbols and red filled symbols are for SOPHIE and SOPHIE+, respectively. The same scale is used for all stars to ease comparison. \emph{Bottom row:} generalised Lomb-Scargle periodogram for the SOPHIE and SOPHIE+ H$\alpha$ index. It was assumed that exists no offset exists between the index measurement of SOPHIE and SOPHIE+. The red vertical dotted lines signal the position of the highest peaks whose periods are annotated, the blue shaded areas represent the 2-$\sigma$ intervals for the stellar rotational periods listed in Table~\ref{table.stellarparams}, and the green vertical dashed lines corresponds to the maximum a posteriori orbital period of the orbiting companions. \label{fig.halpha}}
\end{figure*}

The atmospheric parameters (stellar effective temperature $T_\mathrm{eff}$, surface gravity $\log{(g)}$, and metallicity [Fe/H]) of the four observed targets were computed from the combined SOPHIE spectra. The method is described in \citet{santos2004} and \citet{sousa2008}. Stellar masses $M_\star$ are then derived from the calibration of \citet{torres2010b} with a correction following \citet{santos2013}. Errors were computed from 10 000 random draws from the posterior distribution of the stellar parameters, assumed Gaussian, and free of correlations. The stellar ages are derived by interpolation of the PARSEC \citep{bressan2012} tracks as described by \citet{dasilva2006}. The obtained parameters are listed in Table~\ref{table.stellarparams}. The statistical uncertainty in the stellar masses obtained by this method is known to be underestimated. Instead, we used a conservative 10\% error in all computations requiring the stellar mass, such as the minimum masses of the companions. For HD16175, a number of determinations of the stellar parameters exist in the literature \citep{peek2009, santos2013, sousa2015, jofre2015}. They all agree within the error bars. We decided to retain the determination by \citet{sousa2015} because they used the same method that we employed for the remaining stars.

Using the Hipparcos parallaxes, and the measured apparent magnitudes, we obtained the absolute magnitudes in the V band, $M_V$. With these absolute magnitudes, along with the bolometric correction tabulated by \citet{cox2000} and the self-consistent value adopted for the absolute bolometric magnitude of the Sun, $M_{\mathrm{bol}, \odot} = 4.61$ \citep{torres2010}, we can compute the stellar luminosity in solar units. The obtained values are listed in Table~\ref{table.stellarparams}, and the uncertainties are propagated assuming a negligible error for the bolometric correction. In Fig.~\ref{fig.hr} the position of the target stars in the effective temperature - luminosity plane is compared with the PARSEC theoretical tracks \citep{bressan2012}. We notice that all stars have evolved away from the zero-age main sequence, which is represented by the bottom points of each track. The most evolved star is HD221585, whose age is estimated at $6.2\pm0.5$ Gyr. The remaining stars are around 3 Gyr in age. Their evolution can be quantified using the evolution metric from \citet{wright2005}, $\Delta M_V$, which measures the magnitude difference in the $V$ band between the star and the main sequence. This metric is given in Table~\ref{table.stellarparams} and confirms that HD221585 is the most evolved star in our sample, followed by HD214823. Also, the stellar masses estimated using the \citet{torres2010b} calibration do not generally agree with those that would have been obtained from interpolating the PARSEC tracks. This justifies the use of a 10\% error in the stellar masses for the computation of the derived quantities. A similar discrepancy is seen for the ages of the stars, also hinting at a systematic uncertainty unaccounted for in the error bars listed in Table~\ref{table.stellarparams}.

Finally, we used the SOPHIE CCF to estimate the projected stellar rotational velocities, $v \sin I_*$, using the calibration by \citet{boisse2010}. The value of the projected rotational velocity of the star determined spectroscopically by  \citet{peek2009} for HD16175, $v \sin I_* = 4.8\pm0.5$ \kms\ , is in agreement with the value obtained from the SOPHIE CCF.

\begin{figure}[h]
\center
\input{figures/HRdiagram_withiso.pgf}
\caption{Position of the four target stars in the luminosity-effective temperature plane. The PARSEC theoretical tracks \citep{bressan2012} are plotted using shades of grey for different initial stellar masses labelled in solar units at the bottom of each track. The corresponding isochrones are plotted as dotted lines and labelled in light grey.
\label{fig.hr}}
\end{figure}

\subsection{Activity indexes \label{sect.activity}} 

The classical activity index, based on the Ca II H and K lines, \logR,\  cannot be obtained from the ELODIE spectra because tghe H and K lines are not included in the wavelength range. Also the \logR\ proxy is not easily exploitable from the SOPHIE spectra of these targets due to a poor instrumental response and typically low flux level in this region of the stellar spectrum. As a consequence, the SOPHIE \logR\ time series are usually very noisy. In contrast, the average \logR\ is probably a good indicator of the mean activity level of the star and, therefore, of the expected RV jitter \citep[e.g.][]{santos2000,wright2005} and stellar rotational period \citep{mamajekhillenbrand2008}. The RV jitter level estimated using the prescription of \citet{wright2005} is expected to be of a few \ms, which is comparable with the photon noise error listed in Table~\ref{table.obslog}. The estimated rotational periods are listed in Table~\ref{table.stellarparams}, where the error bars correspond to the dispersion originated in the \logR\ measurements. As expected, stars showing a higher mean value of \logR\ (HD16175 and HD214823) also present higher rotational velocities.

The H$\alpha$ line is formed at a lower altitude in the stellar atmosphere than the Ca II H and K lines, but it was shown to be a good chromospheric indicator \citep[e.g.][]{mauasfalchi94} and is  used frequently \citep[e.g.][]{pasquinipallavicini1991, montes1995, cincunegui2007a, cincunegui2007b, robertson2013, robertson2014, neveu-vanmalle2016}. Furthermore, at the spectral location of the H$\alpha$ line SOPHIE and ELODIE have a much higher instrumental response and, therefore, the activity index based on the H$\alpha$ transition line can be readily computed. The activity index is defined as the ratio between the flux in the H$\alpha$ line and a reference flux near the continuum level. We chose a 10.76 \AA\ window centred at 6550 \AA\ and a 8.75 \AA\ window centred at 6580 \AA\ for the
reference windows. We used 0.68 \AA\ window centred at 6562.808 \AA\footnote{This is narrower than the 1.5-\AA\ window used by \citet{cincunegui2007a} to compute the H$\alpha$ index in medium resolution spectra and also than the 1.6-\AA\ window employed by \citet{gomesdasilva2014}. We believe that such large windows are at the base of the variability in correlation between the H$\alpha$ index and Ca II H \& K index reported by these authors. A thorough analysis is needed.} for the line flux measurement. The H$\alpha$ activity index cannot be used to compare the activity level of different stars, since the flux ratio used in its definition depends not only on activity level, but also on the spectral type of each star. The index can be corrected as carried out by \citet{cincunegui2007a}  by subtracting an estimation of the photospheric flux. In the scope of this paper, we are interested exclusively in the relative activity changes of each star and, therefore, we did not correct the H$\alpha$ index. 

The H$\alpha$ measurements are plotted in the upper row of Fig.~\ref{fig.halpha}. The lower row presents the periodogram computed using only the SOPHIE and SOPHIE+ measurements. The ELODIE measurements were excluded from the frequency analysis because an offset is expected between ELODIE and SOPHIE H$\alpha$ measurements, but has not yet been calibrated. The position of the highest peak in the periodogram is in agreement with the expected stellar rotational period obtained based on the \logR\ indicator (see Table~\ref{table.stellarparams}) for HD214823 and HD221585. For HD16175 and HD191806, power is seen at the expected rotational period ($\sim$20 days), but the highest peak is elsewhere. For HD16175, if we remove the observation with the lowest S/N ratio, 31, compared to the mean of 53, the highest peak shifts to 12.1 days, which is close to the expected rotational period of $19.7\pm 5.5$ days.

The periodogram of the H$\alpha$ time series of HD191806 shows a prominent peak at around 1600 days, similar to the period of the detected companion. However, the correlation between velocities and H$\alpha$ measurements is weak (Pearson's coefficient < 0.3), and when a sinusoidal model with this frequency is fit to the H$\alpha$ data, the amplitude is compatible with zero\footnote{Precisely, the 95\% HDI of the amplitude distribution contains zero.}. Furthermore, we computed the Bayes factor between this model and a model without variation. Using the estimator of \citet{perrakis2014} we found that the simpler model is $86.1\pm0.2$ times more probable than the sinusoidal model. In contrast, when the period of the putative signal is allowed to vary, we obtained a Bayes factor of $2.2\pm0.2$ for the sinusoidal model. This is considered mild evidence against the null hypothesis and "not worth more than a bare mention" \citep{kassraftery1995}. We conclude that the peak observed at this frequency is not significant. Therefore, none of the target stars exhibit activity variations with periods larger than $\sim$1000 days. The activity level of HD214823 seems to be decreasing on a timescale that is much longer than the current time base. Unfortunately, the precision of the \logR\ measurements is not sufficient to detect a similar variability.

\section{Bisector analysis \label{sect.bisector}}
The stellar RV signature of a planetary-mass companion can be mimicked by blended stellar systems and nearly face-on stellar binaries \citep[e.g.][]{santos2002, diaz2012, wright2013}. However, while the reflex motion produced by a planetary-mass companion induces a shift of the stellar spectral lines without changing their shape, blended stellar systems and unresolved companions induce some level of variation because of the presence of an additional set of lines. This variation is reflected, in principle, in the bisector velocity span (BVS) obtained from the CCF. However, it may be too small to be detected, especially when the star is a slow rotator \citep[see][]{santerne2015}. Stellar activity is also known to produce variations in the BVS over the timescale of the rotational period of the star \citep[see e.g.][]{boisse2011} and also throughout stellar activity cycles \citep[see e.g.][]{diaz2016a}.

We searched for significant variations in the SOPHIE BVS \citep{queloz2001}, as well as correlations between the bisector velocity span and RV measurements. The SOPHIE and SOPHIE+ measurements were considered separately, since the instrument upgrade may have induced changes in the instrumental line profile. The results are presented in Table~\ref{table.bisanalysis}, where we list the $p$-value of the $\chi^2$ statistic under the null hypothesis (i.e. no BVS variation) for each star, assuming the photon noise on the BVS is twice that of the RV measurement \citep{santerne2015}. No $p$-value is below the customary limit of 0.05. In view of the excessive tendency of $p$-value analysis to reject the null hypothesis \citep[see e.g.][]{sellke2001}, we conclude that no significant variation is seen in the bisector time series, and that, therefore, the variations seen in the RV time series are produced by orbiting substellar companions.

\begin{table}
\caption{Results of the bisector analysis for SOPHIE and SOPHIE+ data. $\chi^2$ statistics under the null hypothesis of no bisector variation. \label{table.bisanalysis}}
\begin{tabular}{l | cc  cc }
\hline
\hline
Target & \multicolumn{2}{c}{SOPHIE} & \multicolumn{2}{c}{SOPHIE+}\\
        & $\chi^2$ & $p$-value & $\chi^2$ & $p$-value \\
\hline
HD16175                 & -- & -- &34.86 & 0.07 \\
HD191806        & 24.45 & 0.55 & 20.75 & 0.19\\
HD214823        & 14.28 & 0.28 & 4.11 & 0.94\\
HD221585        & 3.303 & 0.51 & 8.27 & 0.98\\
\hline
\end{tabular}

\end{table}

\section{Data analysis \label{sect.analysis}}
\subsection{The model \label{sect.model}}
The stellar RV data of instrument $k$ ($\{v_i\}$ with $i = 1, ..., N_k$)  were modelled as originating from independent normal distributions with mean $f(t_i, \teta)$ and standard deviation $\sigma$. The mean $f(t_i, \teta)$ is the physical model for the RV variations at time $t_i$ (with parameter vector $\teta$), which we describe below. The variance can be written $\sigma^2 = \sigma_i^2 + \sigma_{Ji}^{(k)\,2}$, where $\sigma_i$ is the internal photon noise and calibration error of the RV measurement at time $t_i$ and $\sigma^{(k)}_{J}$ is the additional noise introduced in the model to represent activity jitter and instrument systematics, which could in principle also be a function of time. The term $\sigma_{Ji}^{(k)}$ depends on the instrument under consideration, hence the index $k$. The likelihood for this model then is the product over all of the instruments $\{k\}$ with $k = 1, ..., N_\mathrm{inst}$ and all velocity measurements for each instrument as follows:
\begin{equation}
\mathcal{L}(\teta) = \prod_{k=1}^{N_\mathrm{inst}} \prod_{i=1}^{N_k} \frac{1}{\sqrt{2\pi}\sqrt{\sigma_i^2 + \sigma_{Ji}^{(k)\,2}}} \exp{\left[-\frac{(v_i - f(t_i, \teta) - \gamma^{(k)})^2}{2 (\sigma_i^2 + \sigma_{Ji}^{(k)\,2})}\right]}\; ,
\label{eq.likelihood}
\end{equation}
where the additional parameter $\gamma^{(k)}$ represents the zero-point level, which changes between instruments. In practice we fixed $\gamma$ to zero for one instrument and fitted the relative offset of the remaining instruments, $\delta_\mathrm{RV}^{(k k^\prime)} = \gamma^{(k)} - \gamma^{(k^\prime)}$. When possible, we chose SOPHIE (before the upgrade) as the reference instrument, since there is some prior information on the SOPHIE / ELODIE offset (see below). 

The physical model $f(t_i, \teta)$ at time $t_i$ is
\begin{equation}
f(t_i, \teta) = d_m(t_i, \teta) + \sum_j \kappa_j(t_i, \teta)\;\;, \label{eq.physmodel}
\end{equation}
where $\kappa_j$, with $j \geq 0$, are Keplerian curves representing the reflex motion produced by an orbiting companion to the host star or the magnetic activity effect, usually appearing at the rotational frequency of the star and its harmonics \citep{boisse2011}. The Keplerian curves were parametrised using the orbital period ($P$) and eccentricity ($e$), velocity amplitude ($K$), argument of periastron, ($\omega$) and mean longitude at a given epoch ($L_0$). The term $d_m(t_i, \teta)$ is a $m$th-degree polynomial representing secular accelerations, which are known as stellar drifts. In principle, the physical model could contain an additional term representing the long-term stellar activity effect (i.e. the effect of stellar cycles) as done by \citet{diaz2016a}. As no cycle was detected for any of our stars, this term is not included in the current model description. 

Following the procedure described in \citet{diaz2016a}, $\sigma^{(k)}_J$ was allowed to increase linearly with activity level, measured by the $H_\alpha$ index for SOPHIE data. The value of the additional noise for measurement $i$, $\sigma^{(k)}_{Ji}$ takes the form
\begin{equation}
\sigma^{(k)}_{Ji} = \sigma_J|_\mathrm{min} + \alpha_J \cdot (H\alpha_i - \min(\{H\alpha\}))\;\;,
\label{eq.jittermodel}
 \end{equation}
where $\sigma_J|_\mathrm{min}$ is the additional noise at activity minimum, which includes the systematic instrumental noise, and $\alpha_J$ is the sensitivity of the noise to activity level. This model has two additional parameters: $\sigma_J|_\mathrm{min}$ and $\alpha_J$. In contrast, as ELODIE velocities are more concentrated in time, a constant jitter term was used. 

The parameter priors are presented in Table~\ref{table.priors}. The prior distributions chosen for the zero-point offset between ELODIE and SOPHIE come from the calibration presented by \citet{boisse2012b} using the $B-V$ value listed in Table~\ref{table.stellarparams}. Between SOPHIE and SOPHIE+, we chose a zero-centred normal distribution with a standard deviation of 10 \ms. This comes from preliminary work done on the SOPHIE data, which indicates that once corrected for the seeing effect, the zero-point offset introduced by the upgrade is insignificant.

\subsection{Posterior sampling}
We sampled the posterior distributions of the model parameter using the Markov chain Monte Carlo described in \citet{diaz2014}. We obtained the starting point of the MCMC algorithm  using a genetic algorithm included in the \emph{yorbit} package \citep{segransan2011, bouchy2016}. Ten independent samplers were run for each system for 500'000 iterations and these results were combined as described in \citet{diaz2014}. In this way, we obtained over 20 000 independent posterior samples for each target.

\subsection{Astrometric analysis}
We analysed the available Hipparcos astrometry \citep{hipparcos} of the four targets stars. The new Hipparcos reduction \citep{vanleeuwen2007} was employed to search for signatures of orbital motion in the Intermediate Astrometric Data (IAD). The analysis was performed following \citet{sahlmann2011}, where a detailed description of the method can be found.

Using the parameters of the radial velocity orbit (Table \ref{table.params}), the IAD was fitted with a seven-parameter model, where the free parameters are the inclination $i$, the longitude of the ascending node $\Omega$, the parallax, and offsets to the coordinates and proper motions. A two-dimensional grid in $i$ and $\Omega$ was searched for its global $\chi^2$-minimum. The statistical significance of the derived astrometric orbit is determined with a permutation test employing 1000 pseudo-orbits.  

We did not detect significant orbital signal in the Hipparcos astrometry for any of the four planets. Because none of the orbital periods of the planets is fully covered by Hipparcos measurements, we are unable to set upper limits to the companion masses as we did in the past \citep[e.g.][]{diaz2012}. The only exception is HD221585b, for which Hipparcos observed 94 \% of the orbit and where the non-detection of an astrometric signature excludes a stellar mass for the companion.
 
\section{Results \label{sect.results}}

In this section we discuss the results of the modelling of the RV data. The RV measurements are plotted together with the fitted models in Fig~\ref{fig.rvcurves}. A summary description of the posterior distribution is given in Table~\ref{table.params}, where the posterior mean and the equal-tailed 68.3\% confidence interval is given. We also reported the 95\% highest density interval (HDI) limit when the lower limit of the 95\% HDI was at the limit imposed by the priors listed in Table~\ref{table.priors}.

\subsection{New companions}

\begin{figure}
\input{figures/HD191806_k1_sophiepc_sophienoisemodel1_jitteroffset.pgf}
\input{figures/HD191806_k1d1_sophiepc_sophienoisemodel1_jitteroffset.pgf}
\caption{Posterior distributions of the additional noise terms (upper panels) and instrument offsets (lower panels) for HD191806 under two competing models: with (right column) and without (left column) a secular acceleration term (order $m$ in Eq.~\ref{eq.physmodel}). In the upper panels, we depict the posterior distribution of additional noise at activity minimum for SOPHIE and SOPHIE+ (see Sect.~\ref{sect.model}) and the amplitude of the (constant) additional noise for ELODIE. In the lower panels, the dotted curves correspond to prior distributions.\label{fig.HD191806_jitteroffset}}
\end{figure}

\subsubsection{\object{HD191806}}

The model used to fit the HD191806 RV data consisted of a single Keplerian curve plus a linear trend (i.e. $j=1$ and $m=1$ in Equation~\ref{eq.physmodel}). The presence of a long-term trend is not obvious in the original data set and depends on the values taken by the instrument offsets. However, the model without a secular acceleration ($m=0$) results in a worse fit to the data because its posterior distribution for the SOPHIE / SOPHIE+ offset peaks near -50 \ms, i.e. far from the prior centre (see Fig.~\ref{fig.HD191806_jitteroffset}, left column). In other words, the model without acceleration requires an unrealistically high value of this parameter to fit the data reasonably well. As a consequence, the Bayes factor, which is computed using the techniques of \citet{perrakis2014} and \citet{chibjeliazkov2001} \citep[see also][]{diaz2016a}, shows that the model with the acceleration term is $280\pm75$ times more probable than the simpler model despite the use of additional parameters. Incidentally, under the model without linear drift, the amplitude of the additional noise at activity minimum, $\sigma_J|_\mathrm{min}$, for SOPHIE+ has a posterior distribution similar to that of SOPHIE (Fig.~\ref{fig.HD191806_jitteroffset}). This would only be expected if the activity jitter dominates completely over the instrumental noise, which is not expected for this star. However, this excessive jitter term for SOPHIE+ is not punished in the Bayes factor computation because we chose flat priors for this parameter. When the $m=1$ model is used, the posterior distributions of the jitter term and the offsets are as expected (Fig.~\ref{fig.HD191806_jitteroffset}, right column). We conclude that the velocity of HD191806 exhibits a secular acceleration.

Under the model with a linear trend, the velocity modulation has an amplitude of 140 \ms\ and a period of $1606.3\pm7.2$ days (4.4 years). The frequency spectrum of the H$\alpha$ time series of HD191806 exhibits a peak with the same period as the RV modulation. However, as discussed in Sect.~\ref{sect.activity}, this periodicity is not significant. The data favour a model without activity variation at this period. Furthermore, no significant variation of the bisector span was detected. It seems unlikely that stellar activity is able to produce a 140 \ms\ RV modulation without producing a variation in the line bisectors. We therefore conclude that the RV variation is due to a companion in orbit around HD191806. Nevertheless, we remain cautious concerning the accuracy of our parameter determination, which can potentially be affected by activity. Long-term follow-up of this target should permit us to better understand the H$\alpha$ variability and its effect on the system parameters.

Additionally, we detect a significant eccentricity of $e = 0.259\pm0.017$. The companion around HD191806 has therefore a minimum mass of $8.52\pm0.63$ \MJ. The linear trend has an amplitude of $11.4\pm1.7$ \ms $\text{yr}^{-1}$.

\subsubsection{\object{HD214823}}
The RV data of HD214823 are adequately reproduced by a model with $j=1$ and $m=0$, i.e. a model with a single Keplerian curve, with an amplitude of around 280 \ms and a period of 1877 days (5.1 years). The orbit is fully covered by the SOPHIE and SOPHIE+ data, but the ELODIE measurements help constrain the period. The velocity offset between ELODIE and SOPHIE ($24 \pm 19$ \ms) is in agreement with the value expected from the calibration of \citet{boisse2012b}, which is $57\pm23$ \ms (see Table~\ref{table.priors}), but the precision is improved. In contrast, the offset between SOPHIE and SOPHIE+ is compatible with zero, as expected.

The activity level of HD214823 seems to be decreasing (Fig.~\ref{fig.halpha}). Although some power is seen in the frequency spectrum of the H$\alpha$ index at the period of the Keplerian curve, this is most probably due to long-term activity evolution. The highest peak in the H$\alpha$ periodogram (13.1 days) agrees with the expected rotational period of the star based on the calibration by \citet{mamajekhillenbrand2008}. Additionally, there are no significant signals in the bisector velocity span (Table~\ref{table.bisanalysis}). We conclude therefore that the detected radial velocity variation originates from a companion with a minimum mass $m_c \sin i = 19.4 \pm 1.4$ \MJ. HD214823 b is therefore a brown dwarf candidate on a mildly eccentric orbit, $e = 0.154\pm0.014$. 

\subsubsection{\object{HD221585}}
HD221585 exhibits RV variations with an amplitude of around 28 \ms\ and a period of 1173 days (3.2 years) under the model with no secular acceleration and a single Keplerian curve. The final orbital fit of HD221585 was only possible after the addition of the last few SOPHIE+ measurements. The reason for this is the relatively small number of SOPHIE measurements, which impeded a correct determination of the velocity offset between instruments and allowed for the presence of secular accelerations in a preliminary analysis. Here again, the posterior velocity difference between SOPHIE and SOPHIE+ is compatible with no offset, and the SOPHIE / ELODIE offset agrees with the expected value and is determined with improved precision (see discussion below).

The activity level of HD221585 has remained unchanged for over ten years, as seen in the time series of the H$\alpha$ measurements. However, we interpret the peak at 35-day period in the periodogram of the activity indexes as an indication of the stellar rotational period, which is in agreement with the expectation from the activity level measured in the Ca II H and K lines. No power is seen at periods close to that of the Keplerian curve and, furthermore, the bisector velocity span does not show any significant variability. These facts lead us to conclude that the RV variability is produced by an orbiting companion with minimum mass of $m_c \sin i = 1.61\pm0.14$ \MJ. 

The amplitude of the reflex motion of HD221585 is the smallest among the stars studied here. The eccentricity is therefore determined less precisely and did not lead to a significant detection. The eccentricity 95\% upper limit is 0.24. The companion is at a distance of 2.3 au from its star, and is therefore unlikely that the circularisation of the orbit has occurred through tidal interaction.

\subsection{Orbit update}
\subsubsection{\object{HD16175}}
A companion to HD16175 was discovered by \citet{peek2009} based on RV measurements obtained at Lick Observatory. They reported a companion on an eccentric ($e = 0.59\pm0.11$), 990-day orbit with a minimum mass of 4.40 \MJ. The host star is an evolved, metal-rich G0 subgiant.

We combined the Lick data with the ELODIE and SOPHIE+ data to improve the orbital and physical parameters of the system. We used a constant additional noise term for the Lick data. As we do not have any information on the velocity offset between the Lick and ELODIE or SOPHIE measurements, we used a flat prior spanning 400 \ms\ (see Table~\ref{table.priors}). Fortunately, one Lick measurement overlaps in time with the ELODIE measurements (Fig.~\ref{fig.rvcurves}) and, therefore, the offset is constrained by the data itself with a precision of 2.7 \ms.

Our results are in agreement with those of \citet{peek2009}, but the precision is improved by a factor ranging between 2 and 7 for the orbital parameters. We found a period $P=995.4\pm2.8$ days, and eccentricity $e = 0.637 \pm 0.020$, which is slightly higher than the value reported previously, and a velocity amplitude, $K = 103.5\pm5.0$ \ms. Since no variation in the activity index nor in the bisector is seen, we conclude that this variation is due to a companion with a minimum mass of $m_c \sin i = 4.77\pm0.37$ \MJ. 

\input{tableparams.tex}

\section{Giant planets in the habitable zone \label{sect.hz}}

\begin{figure*}[t]
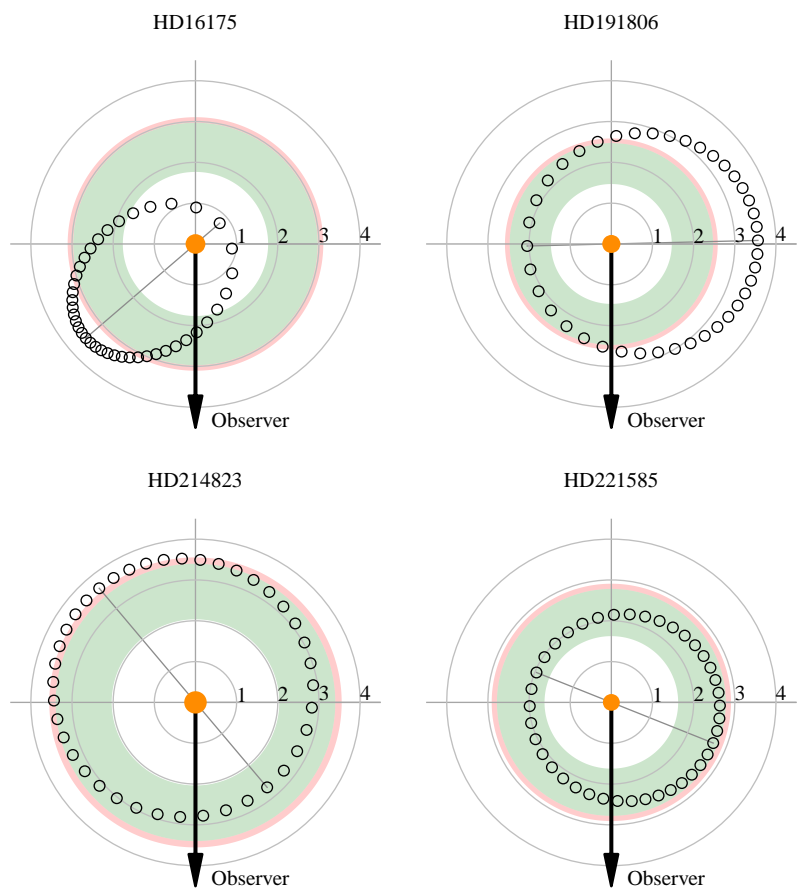

\begin{center}
\input{figures/HD16175_orbit_k1_sophiepc_lick_sophienoisemodel1.pgf}
\input{figures/HD191806_orbit_k1d1_sophiepc_sophienoisemodel1.pgf}

\input{figures/HD214823_orbit_k1_sophiepc_sophienoisemodel1.pgf}
\input{figures/HD221585_orbit_k1_sophiepc_sophienoisemodel1.pgf}
\end{center}

\caption{Schematic view of the systems. The orbital plane is represented. The maximum a posteriori companion orbits are indicated with the empty black points that are equally spaced in time over the orbit. The orbital movement is counter-clockwise and angles are measured counter-clockwise from the negative x-axis. The semi-major axis of the orbit is shown as a thin grey line. The host star is at the centre and the area of the orange circle is proportional to its luminosity. The concentric circles are labelled in astronomical units and the black thick arrow points towards the observer. The filled green area is the habitable zone comprised between the "runaway greenhouse" limit and the "maximum greenhouse" limit, according to the model of \citet{kopparapu2013}. The red area corresponds to the increased habitable zone if the outer edge is defined by the "early Mars" limit. \label{fig.hzorbits}}
\end{figure*}

\begin{table}
\begin{center}
\caption{Habitable zone location and average incident flux. Distances are given in au.\label{table.hz}}
\begin{tabular}{l ccccc}
\hline\hline
          & \multicolumn{2}{c}{Inner edge} & \multicolumn{2}{c}{Outter edge} \\
Target &         RGH & MoGH & MaGH & EMA & $<\mathcal{F}>/\mathcal{F}_\oplus$\\
\hline
HD16175 &1.74 & 1.76  & 2.97 & 3.11 &$0.91\pm 0.10$\\
HD191806&1.44 & 1.47 & 2.47 & 2.58& $0.30\pm 0.03$\\
HD214823&1.99 & 2.03 & 3.40 & 3.55& $0.44\pm 0.07$\\
HD221585&1.61 & 1.62 & 2.78 & 2.91& $0.50\pm 0.05$\\
\hline
\end{tabular}
\tablefoot{RGH: runaway greenhouse; MoGH: moist greenhouse; MaGH: maximum greenhouse; EMA: early Mars.}
\end{center}
\end{table}

The stellar effective temperatures and luminosities computed as described in Sect.~\ref{sect.stellarparams} can be used to estimate the location of the habitable zone (HZ) around each star. This is done using Eq. 2 and 3 of \citet{kopparapu2013} and the erratum \citet{kopparapu2013b} for different habitable zone limits. They are listed in Table~\ref{table.hz}. Following \citet{kopparapu2013} the inner edge of the HZ is taken as the distance where runaway greenhouse effect takes place. Although this is more liberal than choosing the "moist greenhouse" limit for stars hotter than around 5500 K (see their Fig.~8), the candidates presented here are closer to the outer edge of the habitable zone. For the outer edge, we chose the "maximum greenhouse" limit, where the heating effect of the greenhouse is outweighed by the Rayleigh scattering by CO$_2$. The more liberal "early Mars" expands the outer edge only slightly (see Fig.~\ref{fig.hzorbits}). 

A schematic view of the results is provided in Fig.~\ref{fig.hzorbits}. On account of their orbital eccentricity, the majority of the detected companions make excursions outside of the HZ throughout their orbit. The exception is HD221585 b, whose entire orbit is spent within the HZ. In contrast, HD214823 b, spends around 23\% of its orbit outside\ the HZ. The situation is more critical for HD191806 b, which spends 68\% of its orbit outside the HZ. The situation of HD16175 b is even more extreme with 21\% of its orbit spent within the inner edge of the HZ and more than 38\% spent outside the outer edge. However, \citet{williamspollard2002} argued that long-term climate stability is dictated by the mean incident flux throughout the orbit and not the length of time spent in the HZ. The mean incident flux over an orbit with respect to the mean flux received by a planet at 1 au orbiting the Sun ($\mathcal{F}_\oplus$) is
\begin{equation}
\frac{<\mathcal{F}>}{\mathcal{F}_\oplus} = \frac{L}{L_\odot}\frac{1}{a^2 (1 - e^2)^{1/2}}\;\;,
\end{equation}
where $L$ is the stellar luminosity and $a$ is the semi-major axis of the orbit (in au). The average incident flux is listed in Table~\ref{table.hz}.

\begin{figure}
\input{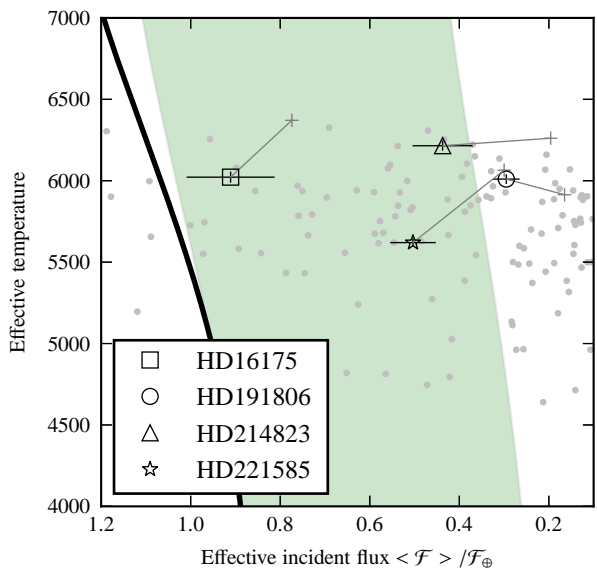}
\caption{Locus of the habitable zone in the effective temperature - incident flux plane. The green filled area is delimited on the right by the maximum greenhouse limit and on the left by the moist greenhouse limit. The solid black curve represents the position of the runaway greenhouse limit. The position of the candidate planets are marked with different symbols, as indicated in the legend. The straight lines connect the current positions of the candidates with the positions they had when their host stars were 1 Gyr old (indicated by plus signs), according to the PARSEC evolution tracks (see text for detail). The grey points are giant planets ($M\sin i > 0.1$ \MJ) orbiting non-evolved stars \citep[$\Delta M_V < 2.0$; ][]{wright2005}.\label{fig.hz}}
\end{figure}

In Fig.~\ref{fig.hz} \citep[inspired by Fig.~8 of ][]{kopparapu2013} we place the resulting average incident fluxes with respect to the limiting HZ fluxes for stars with different effective temperatures $T_\mathrm{eff}$. The filled region is the HZ enclosed by the "moist greenhouse" limit and the "maximum greenhouse" limit, which correspond to the same region indicated in the orbit plots (Fig.~\ref{fig.hzorbits}). The solid black curve is the locus of the runaway greenhouse limit. We see that all candidates but HD191806 b lie well within the HZ of their star. Even HD16175 b, with its very eccentric orbit, receives an average flux that would allow liquid water to exist in the surface of a hypothetical rocky moon provided that the satellite is capable of maintaining a relatively constant temperature throughout the orbit, despite the change in insolation of a factor $[(1 + e)/(1 - e)]^2 \sim 20$.  Hypothetical rocky satellites orbiting HD221585 b and HD214823 b have good prospects for habitability. 

However, as mentioned above, all the target stars evolved at least slightly from their primitive position in the main sequence. Therefore, the positions of their companions in Fig.~\ref{fig.hz} are not representative of the positions during the main-sequence lifetime. Using the PARSEC tracks, we estimated the effective temperatures effective incident fluxes when the stars were 1 Gyr of age. For this, we used the values of the stellar masses and metallicities listed in Table~\ref{table.stellarparams} and assumed that the companion orbits have not evolved. The past positions obtained under this hypothesis are indicated with plus signs in the diagram of Fig.~\ref{fig.hz} and connected to the current positions by straight lines. Because the stars were less luminous in the past and we assumed unchanged orbits, the companions move to the right in the diagram as we move to the past. HD214823 b and HD221585 b were located outside the HZ when the stars were 1 Gyr old; in contrast, the companion around HD16175 is in the "continuously habitable" zone. This computation is, however, extremely dependent on the current stellar parameters (ages and masses) and on the stellar models used to trace the position of the stars into the past.  The evolution shown in Fig.~\ref{fig.hz} does not correspond to the evolution one would guess by following the closest evolution track shown in Fig.~\ref{fig.hr} across the isochrones. This is mainly due  to the discrepancy in the mass determination mentioned above. This underlines the need for accurate stellar masses and ages and the importance of the space mission PLATO \citep{rauer2014}.

\section{Summary and discussion \label{sect.discussion}}
\begin{figure}
\input{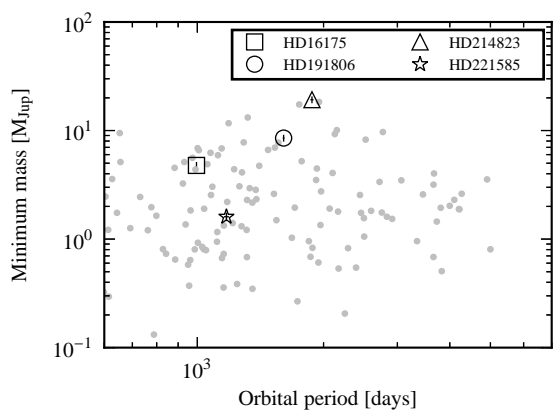}
\caption{Position of the new companions in the mass-period diagram. The grey points are giant planets ($M\sin i > 0.1$ \MJ) reported in the literature \citep{han2014} with orbital period $P > 500$ days orbiting non-evolved stars \citep[$\Delta M_V < 2.0$; ][]{wright2005}. Note that the error bars for our detections are smaller than the symbols.\label{fig.context}}
\end{figure}

We detected three new companions to solar-type stars on long-period orbits and an updated orbit for the Jupiter-mass candidate HD16175 b, which was first reported by \citet{peek2009}. The position of the companions in the mass-period diagram is not atypical (Fig.~\ref{fig.context}). HD214823 b, with a minimum mass of 19 \MJ\ , is a brown dwarf candidate that bears resemblance to HD168443 c \citep{marcy2001} and HD131664 b \citep{moutou2009b}. The host stars are slightly more massive than the Sun and seem to have started evolving away from the main sequence. The most evolved star studied here is HD221585. An interpolation using the PARSEC tracks \citep{bressan2012, dasilva2006} gives an age of 6.6 Gyr and a radius of 1.7 $R_\odot$. All four stars are metal rich with metallicities, [Fe/H], ranging between +0.17 and +0.30.

We combined data from the ELODIE, SOPHIE, and SOPHIE+ spectrographs along with observations from the Lick planet search. We took a Bayesian approach to model the radial velocity variations to include a priori knowledge on the instrument velocity offsets, whenever it was available. The case of HD191806 illustrates the importance of including this information in the analysis. Indeed, had the SOPHIE / SOPHIE+ offset been left free, the model without an acceleration term would have resulted in a good fit, albeit with an unrealistically high value for this parameter. 

To account for the noise induced by stellar activity and instrument systematics, the model included an additional white noise term.  For SOPHIE data, the amplitude of the additional noise term was allowed to vary linearly with activity level. For SOPHIE+, the 95\% HDI of the noise amplitude at activity minimum includes zero, which means that no additional noise is actually measured. This is probably because of the relatively low precision of the SP2 measurements (see Sect.~\ref{sect.observations}); the scatter of the SOPHIE+ measurements is dominated by photon noise. HD221585 is the least active star in our sample (<\logR> = -4. 86) and exhibits the lowest base-level jitter for SOPHIE+: only $2.1\pm1.4$ \ms\ (upper limit 4.5 \ms). On the contrary, for SOPHIE data taken before the upgrade, noise level in excess of the photon noise was detected. The only exception is HD221585, which was observed only five times before the upgrade. Additionally, for all stars except for HD16705, the most active of the sample, the slope of the error model is compatible with zero at the 95\% level. This means that the additional noise term is compatible with a constant jitter.  The precision improvement between SOPHIE and SOPHIE+, which is evident from the residual plots in Fig.~\ref{fig.rvcurves}, can be quantified using the posterior distributions of the additional noise term: the mean value is lower by a factor of around 3.6. As expected, the dispersion of the residuals is positively correlated with the mean activity level of the stars, <\logR>. 

One attractive feature of the Bayesian analysis performed here is that it teaches us about the instruments in use. For example, the offset between ELODIE and SOPHIE was calibrated by \citet{boisse2012b} based on a sample of around 200 stars. They obtained a relationship between the offset and colour index $B-V$ of the stars, which we used as prior information for our analysis. The posterior distributions of the SOPHIE / ELODIE offsets are in all cases narrower than the prior distribution, i.e. the current data constrain the offset better than the simple calibration. For example, in the case of HD221585, the dispersion of the posterior distribution is only 5.5 \ms compared to the 23 \ms\ used for the prior \citep[Table~\ref{table.priors}; see also][]{boisse2012b}. This is expected because the \citet{boisse2012b} calibration exhibits a dispersion in excess of the typical error of the individual measurements used to derive it. Performing the type of analysis we presented here on a large number of stars could serve as the basis for an improved calibration (for example, allowing for a dependence on stellar metallicity). Similarly, we see that when the SOPHIE data are corrected as described in Sect.~\ref{sect.observations} the velocity difference between SOPHIE and SOPHIE+ is compatible with zero. This is in agreement with the chosen prior information. However, there is a hint of a dependence on colour index $B-V$; the offset posterior distribution for HD221585, the reddest star in our sample, peaks at higher values than for HD191806 and HD214823. To verify this dependence an analysis on a larger number of stars is warranted.

Three of the companions studied are currently in the habitable zone of their host star, which would make hypothetical rocky satellites orbiting around them a suitable place for life to develop. The best technique to find satellites of giant planets is probably the transit method \citep[e.g.][]{kipping2012}, however, the times of inferior conjunction of these companions are only known to a precision of between 9 and 30 days (see Table~\ref{table.params}). Therefore, a ground-based search is likely unfeasible. On the other hand, these stars will certainly become prime targets for the follow-up space mission CHEOPS \citep{broeg2013} that will search for transits of these giant planets and of potentially habitable exomoons \citep{simon2015}. An issue arises concerning the time the candidate has spent in the habitable zone during its lifetime. Indeed, all the stars studied here are slightly evolved, which means that the environment at the candidate orbital distance has changed in the recent past. Out of the four candidates reported here, only HD16175 b would be, according to estimates based on the PARSEC stellar evolution model and the measurement of the stellar masses and metallicities in the so-called "continuously habitable" zone. The remaining candidates were not inside the habitable zone during their early main-sequence lifetimes. However, this conclusion is strongly dependent on the current age of the system, which is presently very uncertain. This stresses the importance of the space mission PLATO \citep{rauer2014} and the accurate determination of the masses and ages of exoplanet hosts.

\begin{acknowledgements}
We thank all the staff of Haute-Provence Observatory for their support at the 1.93-m telescope and on SOPHIE. 

This work has been carried out within the frame of the National Centre for Competence in Research "PlanetS" supported by the Swiss National Science Foundation (SNSF).

JR acknowledges support from the CONICYT/Becas Chile 72140583.

AS and NCS acknowledge the support from Funda\c{c}\~ao para a
Ci\^encia e a Tecnologia (FCT, Portugal) in the form of grants reference
UID/FIS/04434/2013 (POCI-01-0145-FEDER-007672) and PTDC/FIS-AST/1526/2014.
NCS was also supported by FCT through the Investigador FCT contract reference IF/00169/2012
and POPH/FSE (EC) by FEDER funding through the programme "Programa Operacional de Factores de Competitividade - COMPETE. AS is supported by the European Union under a Marie Curie Intra-European Fellowship for Career Development with reference FP7-PEOPLE-2013-IEF, number 627202.

P.A.W acknowledges the support of the French Agence Nationale de la Recherche (ANR), under programme
ANR-12-BS05-0012 "Exo-Atmos". 

This research has made use of the SIMBAD database and of the VizieR catalogue access tool operated at CDS, France. This research has made use of the Exoplanet Orbit Database
and the Exoplanet Data Explorer at exoplanets.org.
\end{acknowledgements}

\input{sophie11_vArXiv.bbl}
\onecolumn
\begin{longtab}
\input{HD16175_rvtable.tex}
\input{HD191806_rvtable.tex}
\input{HD214823_rvtable.tex}
\input{HD221585_rvtable.tex}
\end{longtab}

\begin{appendix}
\section{Prior distributions}
\input{priors.tex}
\end{appendix}

\end{document}

%% file: StellarParams_sophie11.tex
\begin{table*}[t!]
\center
\caption{\label{table.stellarparams}
Stellar parameters.}
\begin{tabular}{llcccc}
\hline\hline
Parameters    		&   		&\object{HD16175} &\object{HD191806}&\object{HD214823}&\object{HD221585} \\
\hline                                   
Right Ascension	&(J2000)	&02:37:01.9	&20:09:28.3	&22:40:19.9	&23:32 54.0\\
Declination		&(J2000)	&+42:03:45.48	&+52:16:34.79	&+31:47:15.33	&+63:09:19.73\\
Spectral Type$^{(1)}$  		&     		& G0			&--   		   & G0V   &	G8IV  \\
V$^{(2)}$              	&     		&$7.291\pm0.011$	&$8.093\pm0.014$  &$8.078\pm0.013$  &$7.474\pm0.011$	\\
$B-V^{(1)}$           	&     		& 0.66		&0.64  		   & 0.63  &   	0.77\\
${M_V}^{\bullet}$		&		&$3.48 \pm 0.08$	&$3.89 \pm 0.08$	&$3.13 \pm 0.14$ &$3.68 \pm 0.08$\\
$\pi^{(3)}$            	&[mas]	&$17.28 \pm 0.67$	&$14.41\pm0.50$     & $10.25\pm0.68$ & $17.40\pm0.60$\\
${L/L_\odot}^{\bullet}$		&		&$3.22 \pm 0.25$	&$2.23 \pm 0.16$	&$4.35 \pm 0.58$	&$2.64 \pm 0.18$\\
${\Delta M_V}^{\bullet, (4)}$	&		&$1.38 \pm 0.08$	&$0.84 \pm 0.08$	&$1.53 \pm 0.14$	&$1.83 \pm 0.08$\\
\hline
$T_{\rm eff}$	& [K] 	&$6022 \pm 34^{(5)}$   &$6010 \pm 30^\bullet$     &$6215 \pm 30^\bullet$	&$5620 \pm 27^\bullet$\\
$\mathrm{[Fe/H]}$		&[dex]	&$+0.37\pm0.03^{(5)}$ &$+0.30 \pm 0.02^\bullet$  &$+0.17 \pm 0.02^\bullet$&$+0.29 \pm 0.02^\bullet$\\
$\log{(g)}$	& [cgs]	        &$4.21 \pm 0.06^{(5)}$ &$4.45 \pm 0.03^\bullet$   &$4.05 \pm 0.10^\bullet$	&$4.05 \pm 0.04^\bullet$\\
<\logR>$^{\bullet, (6)}$		&[dex]	&$-4.77\pm0.15$	&$-4.85\pm0.21$&$-4.79\pm0.14$&$-4.86\pm0.13$\\
${P_\mathrm{rot}}^{\bullet, (7)}$	&[days]	& $19.7 \pm 5.5$ & $20.6 \pm 6.9$ &$17.5 \pm 4.5$ & $33.3 \pm 7.2$\\
${v \sin I_*}^{\bullet, (8)}$&[\kms]		&$5.1$&$3.3$&$5.7$&$3.7$\\
$M_{\star}$ &[M$_{\odot}$]&$1.34\pm0.14^{(5)}$	& $1.14\pm0.12^\bullet$     & $1.22\pm0.13^\bullet$ & $1.19\pm0.12^\bullet$	 \\
Age$^{\bullet, (9)}$		   &[Gyr]		&$3.3\pm0.5$		&$2.9\pm0.4$		&$2.7\pm0.2$		&$6.2\pm0.5$\\
\hline
\end{tabular}
\tablebib{
  $\bullet$ This work; (1) Hipparcos catalogue~\citep{hipparcos}; (2)~Tycho catalogue \citep{hog2000}; (3)~\citet{vanleeuwen2007}; (4)~\citet{wright2005}; the errorbar does not include uncertainties in the parametrisation of the main sequence;  (5)~\citet{sousa2015}; (6)~Mean value and standard deviation obtained from SOPHIE spectra; (7)~Using the calibration by \citet{mamajekhillenbrand2008}; (8)~Computed from the SOPHIE CCF using the calibration by \citet{boisse2010}; uncertainty estimated to be 1.0 \kms; {\bf (9)~Obtained by  interpolation of the PARSEC stellar evolution tracks using the method described by \citet{dasilva2006}.}
}
\end{table*}

%% file: tableparams.tex
\begin{table*}[t]
{\small\center
\caption{Parameter posteriors mean and uncertainties.\label{table.params}}            
\begin{tabular}{l l c c}        
\hline\hline                 
\noalign{\smallskip}

\multicolumn{2}{c}{ Orbital parameters } 		&HD191806	&HD214823 \\
\hline
\noalign{\smallskip}
Orbital period, $P^{\bullet}$ 	&[day]	 	&$1606.3 \pm 7.2$   &$1877\pm15$\\
Orbital period, $P$ 			&[year]	 	&$4.398 \pm 0.020$ &$5.138\pm0.042$\\
RV amplitude, $K^{\bullet}$ 	&[\ms]		&$140.5\pm2.1$ 	&$281.4\pm3.7$\\
Eccentricity, $e$$^{\bullet}$    	& 			&$0.259\pm0.017$ &$0.154\pm0.014$ \\
Argument of periastron, $\omega^{\bullet}$&[deg] &$4.0\pm4.2$	&$125.9\pm6.7$\\
Time of periastron passage, $T_p$	&[BJD - 2'455'000]	&$14\pm18$	&$652 \pm 34$\\
Time of inferior conjunction, $T_c$	&[BJD - 2'455'000]	&$269\pm13$	&$513.6 \pm 9.1$\\
$e^{1/2} \cos(\omega)$		&			&$0.506\pm0.018$	&$-0.228\pm0.035$\\
$e^{1/2} \sin(\omega)$		&			&$0.035\pm0.037$	&$0.315\pm0.034$\\
Mean longitude at epoch, $L_0^{\bullet}$ &[deg] &$36.5\pm1.8$ 	&$144.1\pm 1.2$\\
Epoch				& [BJD - 2'455'000]	&158.6766		&747.1753\\
Systemic velocity, $\gamma^{\mathrm{(SOPHIE)}\bullet}$		&[\kms]	&$-15.3597 \pm 0.0030$ &$-44.5038 \pm 0.0060$\\
RV offset, $\delta_\mathrm{RV}^{\mathrm{(ELODIE\;SOPHIE)}\bullet}$&[\ms]	&$100\pm11$	&$24\pm19$\\
RV offset, $\delta_\mathrm{RV}^{\mathrm{(SOPHIE+\;SOPHIE)}\bullet}$&[\ms]	&$-8.4\pm7.6$ &$-4.8\pm6.2$\\
Linear drift$^\bullet$					&[\ms/yr]					&$11.4\pm1.7$			&--\\
Semi-major axis of relative orbit, $a$		&[AU] &$2.80\pm0.10$ &$3.18\pm0.12$\\
Minimum mass, $M \sin i$ 			&[\MJ] &$8.52\pm0.63$ &$19.2 \pm 1.4$\\
\noalign{\smallskip}
\multicolumn{2}{c}{Noise model$\ddag$} 		\\
\hline
\noalign{\smallskip}
SOPHIE+ noise at activity minimum, $\sigma_J|_\mathrm{min}^{\mathrm{(SOPHIE+)}\bullet}$	&[\ms]	&$3.0\pm 2.2$ [7.6]	&$4.6\pm 3.6$ [12.3]\\
SOPHIE noise at activity minimum, $\sigma_J|_\mathrm{min}^{\mathrm{(SOPHIE)}\bullet}$	&[\ms]	&$11.3\pm3.1$	&$16.0\pm5.8$\\
Slope, $\alpha_J^{\bullet}$						&[\kms]								&$0.69 \pm0.48$ [1.60]&$1.08\pm0.86$ [3.03]\\
ELODIE additional noise, 	$\sigma_J^{\mathrm{(ELODIE)}\bullet}$						&[\ms]	&$14\pm11$ [37]	&$39\pm25$ [96]\\
ELODIE $\mathrm{rms}(\mathrm{O-C})$				&[\ms]		&$14.7\pm1.8$		&$21.1\pm2.6$\\
SOPHIE $\mathrm{rms}(\mathrm{O-C})$				&[\ms]		&$14.62\pm0.55$	&$17.6\pm2.8$\\
SOPHIE+ $\mathrm{rms}(\mathrm{O-C})$				&[\ms]		&$6.07\pm0.84$	&$7.5\pm2.1$\\
\hline

\noalign{\bigskip}
\hline
\hline
\noalign{\smallskip}
\multicolumn{2}{c}{ Orbital parameters } 		&HD221585		&HD16175\\
\hline
\noalign{\smallskip}
Orbital period, $P^{\bullet}$ 	&[day]	 	&$1173\pm16$ 		&$995.4\pm2.8$	\\
Orbital period, $P^{\bullet}$ 	&[year]	 	&$3.212\pm0.044$ 	&$2.7251\pm0.0077$	\\
RV amplitude, $K^{\bullet}$ 	&[\ms]		&$27.9\pm1.6$ 	&$103.5\pm5.0$		\\
Eccentricity, $e^{\bullet}$    	& 			&$0.123\pm0.069$ [0.24]	&$0.637\pm0.020$ \\
Argument of periastron, $\omega^{\bullet}$&[deg] &$-6.8\pm28\dag$	&$221.5\pm2.2$\\
Time of periastron passage, $T_p$	&[BJD - 2'455'000]	&$50\pm93\dag$	&$801.4 \pm 2.6$\\
Time of inferior conjunction, $T_c$	&[BJD - 2'455'000]	&$328\pm28$		&$649 \pm 14$\\
$e^{1/2} \cos(\omega)$		&			&$0.29\pm0.12$	&$-0.597 \pm0.025$\\
$e^{1/2} \sin(\omega)$		&			&$-0.05\pm0.14$	&$-0.528 \pm 0.021$	\\
Mean longitude at epoch, $L_0^{\bullet}$ &[deg] &$338.2\pm3.1$ 	&$42.6 \pm 1.8$	\\
Epoch			& [BJD - 2'455'000]		&1178.7064		&1302.4068\\
Systemic velocity, $V_0^{\bullet}$		&[\kms]	&$6.4332 \pm 0.0046$&	$21.8348 \pm0.0023$\\
RV offset, $\delta_\mathrm{RV}^{\mathrm{(ELODIE\;SOPHIE)}\bullet}$&[\ms]	&$91.2\pm5.5$	&---\\
RV offset, $\delta_\mathrm{RV}^{\mathrm{(SOPHIE+\;SOPHIE)}\bullet}$&[\ms]	&$1.7\pm4.6$ &---\\
RV offset, $\delta_\mathrm{RV}^{\mathrm{(ELODIE\;SOPHIE+)}\bullet}$&[\ms]	&---&$72.8\pm8.3$\\
RV offset, $\delta_\mathrm{RV}^{\mathrm{(LICK\;SOPHIE+)}\bullet}$	&[\kms]	&--&$21.7935\pm0.0027$\\
Semi-major axis of relative orbit, $a$		&[AU] 	&$2.306\pm0.081$ &$2.148\pm0.076$\\
Minimum mass, $M \sin i$ 			&[\MJ] 	&$1.61\pm0.14$ &$4.77\pm0.37$\\
\noalign{\smallskip}
\multicolumn{2}{c}{Noise model$\ddag$} 		\\
\hline
\noalign{\smallskip}
SOPHIE+ noise at activity minimum, $\sigma_J|_\mathrm{min}^{\mathrm{(SOPHIE+)}\bullet}$	&[\ms]	&$2.1\pm 1.4$ [4.5]	&$3.4\pm2.7$ [8.6]\\
SOPHIE noise at activity minimum, $\sigma_J|_\mathrm{min}^{\mathrm{(SOPHIE)}\bullet}$	&[\ms]	&$7.8\pm5.1$ [19.8]	&---\\
Slope, $\alpha_J^{\bullet}$						&[\kms]								&$0.66 \pm0.48$ [1.64] &$0.89\pm0.41$\\
ELODIE additional noise, 	$\sigma_J^{\mathrm{(ELODIE)}\bullet}$						&[\ms]	&$7.5\pm5.6$ [19.2]	&$14\pm12$ [47.6]\\
LICK additional noise, $\sigma_J^{\mathrm{(LICK)}\bullet}$						&[\ms]		&---			&$6.9\pm1.4$\\
LICK $\mathrm{rms}(\mathrm{O-C})$				&[\ms]		&	---			&$9.07\pm0.22$\\
ELODIE $\mathrm{rms}(\mathrm{O-C})$				&[\ms]		&$10.6\pm2.4$		&$2.27\pm0.04$\\
SOPHIE $\mathrm{rms}(\mathrm{O-C})$				&[\ms]		&$6.7\pm1.3$		&	---		\\
SOPHIE+ $\mathrm{rms}(\mathrm{O-C})$				&[\ms]		&$4.04\pm0.40$	&$10.4\pm1.0$\\
\hline
\end{tabular}
\tablefoot{Quoted uncertainties correspond to the 68.3\% equal-tailed confidence interval. Where the 95\%-Highest Density Interval (HDI; defined as the interval containing 95\% of the posterior mass such that the any point inside the interval has density higher than any point outside of it) includes the inferior prior limit, the upper 95\% HDI limit is quoted within square brackets.

$\bullet$: MCMC jump parameter.

$\dag$: Parameter essentially unconstrained for this nearly-circular orbit.

$\ddag$:  For SOPHIE and SOPHIE+ the additional noise for measurement $i$ is $\sigma_{Ji} = \sigma_J|_\mathrm{min} + \alpha_J \cdot (H\alpha_i - \min(\{H\alpha\}))$ (see Eq.~\ref{eq.jittermodel}).
}
}
\end{table*}

%% file: HD16175_rvtable.tex
\begin{longtable}{r r r r r r l}
\caption{Radial velocity measurements of HD16175. \label{table.rvHD16175}}\\
\hline\hline
\multicolumn{1}{c}{BJD} &	\multicolumn{1}{c}{RV} &	\multicolumn{1}{c}{$\sigma_\text{RV}$} &	\multicolumn{1}{c}{BIS} &	\multicolumn{1}{c}{H$\alpha$} &	\multicolumn{1}{c}{$\sigma_{\text{H}\alpha}$}&	\multicolumn{1}{l}{Instrument}\\
-2 453 000  &(\kms) & (\kms) & (\ms) & & \\
\hline
\noalign{\smallskip}
\endfirsthead
\caption{Continued.}\\
\hline
\multicolumn{1}{c}{BJD} &	\multicolumn{1}{c}{RV} &	\multicolumn{1}{c}{$\sigma_\text{RV}$} &	\multicolumn{1}{c}{BIS} &	\multicolumn{1}{c}{H$\alpha$} &	\multicolumn{1}{c}{$\sigma_{\text{H}\alpha}$}&	\multicolumn{1}{l}{Instrument}\\
-2 453 000  &(\kms) & (\kms) & (\ms) & (\kms) & & \\
\hline
\noalign{\smallskip}
\endhead
\hline
\endfoot
\hline
\endlastfoot
 281.5892 &	21.7943 &	0.0079 &	 +0.00 &	0.1120 &	\multicolumn{1}{c}{---} &	ELODIE \\
 332.4859 &	21.7839 &	0.0111 &	 +0.00 &	0.1080 &	\multicolumn{1}{c}{---} &	ELODIE \\
 334.4619 &	21.7889 &	0.0080 &	 +0.00 &	0.1130 &	\multicolumn{1}{c}{---} &	ELODIE \\
2806.6450 &	21.6988 &	0.0047 &	 -6.30 &	0.1173 &	0.0018 &	SOPHIE+ \\
2811.5974 &	21.7307 &	0.0077 &	-22.30 &	0.1171 &	0.0029 &	SOPHIE+ \\
2817.5225 &	21.7641 &	0.0046 &	+14.80 &	0.1121 &	0.0016 &	SOPHIE+ \\
2818.4970 &	21.7572 &	0.0046 &	-29.70 &	0.1133 &	0.0017 &	SOPHIE+ \\
2827.6225 &	21.7897 &	0.0046 &	 -2.50 &	0.1123 &	0.0017 &	SOPHIE+ \\
2835.5595 &	21.8074 &	0.0027 &	+10.80 &	0.1080 &	0.0009 &	SOPHIE+ \\
2842.5673 &	21.8141 &	0.0047 &	 -5.80 &	0.1130 &	0.0017 &	SOPHIE+ \\
2875.5144 &	21.8581 &	0.0045 &	-18.50 &	0.1040 &	0.0016 &	SOPHIE+ \\
2877.4684 &	21.8733 &	0.0047 &	-14.50 &	0.1109 &	0.0017 &	SOPHIE+ \\
2903.2755 &	21.8845 &	0.0039 &	 -6.00 &	0.1101 &	0.0013 &	SOPHIE+ \\
2968.2855 &	21.9006 &	0.0049 &	 -2.70 &	0.1068 &	0.0016 &	SOPHIE+ \\
3149.6295 &	21.8830 &	0.0022 &	 +5.20 &	0.1082 &	0.0007 &	SOPHIE+ \\
3263.6168 &	21.8715 &	0.0048 &	+30.00 &	0.1107 &	0.0016 &	SOPHIE+ \\
3297.3102 &	21.8709 &	0.0051 &	 +6.00 &	0.1085 &	0.0018 &	SOPHIE+ \\
3317.3927 &	21.8547 &	0.0072 &	-20.50 &	0.1163 &	0.0025 &	SOPHIE+ \\
3583.5268 &	21.8179 &	0.0047 &	 +4.80 &	0.1089 &	0.0017 &	SOPHIE+ \\
3696.2663 &	21.7496 &	0.0035 &	+12.20 &	0.1090 &	0.0012 &	SOPHIE+ \\
3871.5878 &	21.8903 &	0.0043 &	 -8.20 &	0.1143 &	0.0014 &	SOPHIE+ \\
3887.5898 &	21.8765 &	0.0027 &	 +5.50 &	0.1103 &	0.0009 &	SOPHIE+ \\
3907.6589 &	21.8667 &	0.0042 &	 -0.30 &	0.1144 &	0.0016 &	SOPHIE+ \\
4000.4483 &	21.8859 &	0.0042 &	+11.80 &	0.1072 &	0.0015 &	SOPHIE+ \\
4002.4649 &	21.8812 &	0.0042 &	-11.30 &	0.1100 &	0.0014 &	SOPHIE+ \\
4076.2874 &	21.8813 &	0.0047 &	+10.70 &	0.1067 &	0.0016 &	SOPHIE+ \\
4089.2898 &	21.8796 &	0.0038 &	 -2.00 &	0.1101 &	0.0012 &	SOPHIE+ \\
4268.5851 &	21.8485 &	0.0042 &	 +7.80 &	0.1124 &	0.0016 &	SOPHIE+ \\
\end{longtable}

%% file: HD191806_rvtable.tex
\begin{longtable}{r r r r r r l}
\caption{Radial velocity measurements of HD191806. \label{table.rvHD191806}}\\
\hline\hline
\multicolumn{1}{c}{BJD} &	\multicolumn{1}{c}{RV} &	\multicolumn{1}{c}{$\sigma_\text{RV}$} &	\multicolumn{1}{c}{BIS} &	\multicolumn{1}{c}{H$\alpha$} &	\multicolumn{1}{c}{$\sigma_{\text{H}\alpha}$}&	\multicolumn{1}{l}{Instrument}\\
-2 453 000  &(\kms) & (\kms) & (\ms) & & \\
\hline
\noalign{\smallskip}
\endfirsthead
\caption{Continued.}\\
\hline
\multicolumn{1}{c}{BJD} &	\multicolumn{1}{c}{RV} &	\multicolumn{1}{c}{$\sigma_\text{RV}$} &	\multicolumn{1}{c}{BIS} &	\multicolumn{1}{c}{H$\alpha$} &	\multicolumn{1}{c}{$\sigma_{\text{H}\alpha}$}&	\multicolumn{1}{l}{Instrument}\\
-2 453 000  &(\kms) & (\kms) & (\ms) & (\kms) & & \\
\hline
\noalign{\smallskip}
\endhead
\hline
\endfoot
\hline
\endlastfoot
 275.3780 &	-15.3677 &	0.0117 &	 +0.00 &	0.1130 &	\multicolumn{1}{c}{---} &	ELODIE \\
 277.3547 &	-15.3837 &	0.0091 &	 +0.00 &	0.1190 &	\multicolumn{1}{c}{---} &	ELODIE \\
 549.5873 &	-15.3973 &	0.0093 &	 +0.00 &	0.1120 &	\multicolumn{1}{c}{---} &	ELODIE \\
 934.4769 &	-15.5600 &	0.0124 &	 +0.00 &	0.1090 &	\multicolumn{1}{c}{---} &	ELODIE \\
 936.4564 &	-15.5960 &	0.0140 &	 +0.00 &	0.1100 &	\multicolumn{1}{c}{---} &	ELODIE \\
 960.4783 &	-15.5660 &	0.0207 &	 +0.00 &	0.1120 &	\multicolumn{1}{c}{---} &	ELODIE \\
1306.4979 &	-15.4853 &	0.0048 &	+69.30 &	0.1077 &	0.0018 &	SOPHIE \\
1307.5544 &	-15.4802 &	0.0034 &	+38.80 &	0.1107 &	0.0012 &	SOPHIE \\
1308.4603 &	-15.5049 &	0.0032 &	+39.80 &	0.1100 &	0.0012 &	SOPHIE \\
1309.4760 &	-15.5015 &	0.0030 &	+30.50 &	0.1094 &	0.0011 &	SOPHIE \\
1310.4984 &	-15.4896 &	0.0039 &	+33.50 &	0.1081 &	0.0014 &	SOPHIE \\
1311.5065 &	-15.4723 &	0.0031 &	+31.20 &	0.1121 &	0.0011 &	SOPHIE \\
1312.4696 &	-15.4838 &	0.0035 &	+45.20 &	0.1099 &	0.0013 &	SOPHIE \\
1313.5716 &	-15.4811 &	0.0029 &	+36.50 &	0.1093 &	0.0010 &	SOPHIE \\
1314.4775 &	-15.4879 &	0.0034 &	+34.50 &	0.1101 &	0.0012 &	SOPHIE \\
1328.5718 &	-15.4754 &	0.0044 &	+22.80 &	0.1108 &	0.0014 &	SOPHIE \\
1330.4228 &	-15.4815 &	0.0049 &	+48.20 &	0.1099 &	0.0020 &	SOPHIE \\
1338.4397 &	-15.4890 &	0.0033 &	+40.50 &	0.1128 &	0.0012 &	SOPHIE \\
1350.5303 &	-15.4741 &	0.0027 &	+41.80 &	0.1081 &	0.0009 &	SOPHIE \\
1352.3593 &	-15.4681 &	0.0025 &	+43.80 &	0.1087 &	0.0009 &	SOPHIE \\
1430.2529 &	-15.4558 &	0.0025 &	+35.70 &	0.1068 &	0.0009 &	SOPHIE \\
2041.4830 &	-15.2115 &	0.0044 &	+32.70 &	0.1063 &	0.0016 &	SOPHIE \\
2042.3868 &	-15.1920 &	0.0045 &	+26.20 &	0.1068 &	0.0015 &	SOPHIE \\
2042.5606 &	-15.1949 &	0.0046 &	+26.30 &	0.1098 &	0.0017 &	SOPHIE \\
2092.3896 &	-15.2056 &	0.0042 &	+33.80 &	0.1093 &	0.0015 &	SOPHIE \\
2401.4313 &	-15.3552 &	0.0047 &	+25.50 &	0.1113 &	0.0018 &	SOPHIE \\
2403.4084 &	-15.4023 &	0.0046 &	+30.70 &	0.1138 &	0.0017 &	SOPHIE \\
2428.4663 &	-15.3728 &	0.0041 &	+46.30 &	0.1120 &	0.0015 &	SOPHIE \\
2435.3729 &	-15.3948 &	0.0047 &	+28.70 &	0.1098 &	0.0018 &	SOPHIE \\
2443.4975 &	-15.3994 &	0.0034 &	+43.30 &	0.1115 &	0.0012 &	SOPHIE \\
2495.4067 &	-15.3919 &	0.0047 &	+52.70 &	0.1111 &	0.0017 &	SOPHIE \\
2527.2413 &	-15.4542 &	0.0083 &	+11.70 &	0.1132 &	0.0033 &	SOPHIE \\
2659.6502 &	-15.4690 &	0.0046 &	+45.70 &	0.1141 &	0.0017 &	SOPHIE \\
2807.3832 &	-15.4279 &	0.0087 &	+61.30 &	0.1203 &	0.0033 &	SOPHIE+ \\
2820.3930 &	-15.4381 &	0.0040 &	+15.20 &	0.1108 &	0.0015 &	SOPHIE+ \\
2877.3723 &	-15.4226 &	0.0047 &	+17.50 &	0.1085 &	0.0016 &	SOPHIE+ \\
2994.6887 &	-15.4092 &	0.0055 &	 +2.80 &	0.1151 &	0.0020 &	SOPHIE+ \\
3106.5175 &	-15.3854 &	0.0034 &	+38.00 &	0.1077 &	0.0012 &	SOPHIE+ \\
3155.4027 &	-15.3637 &	0.0047 &	+30.70 &	0.1102 &	0.0017 &	SOPHIE+ \\
3218.4269 &	-15.3382 &	0.0047 &	+39.20 &	0.1081 &	0.0017 &	SOPHIE+ \\
3426.5652 &	-15.2181 &	0.0024 &	+36.00 &	0.1065 &	0.0008 &	SOPHIE+ \\
3486.5425 &	-15.1750 &	0.0047 &	+18.50 &	0.1116 &	0.0016 &	SOPHIE+ \\
3612.2511 &	-15.1229 &	0.0054 &	 +8.50 &	0.1129 &	0.0019 &	SOPHIE+ \\
3622.2372 &	-15.1220 &	0.0046 &	+38.50 &	0.1077 &	0.0016 &	SOPHIE+ \\
3643.2640 &	-15.1354 &	0.0037 &	+37.80 &	0.1101 &	0.0013 &	SOPHIE+ \\
3793.6204 &	-15.2053 &	0.0043 &	+20.50 &	0.1146 &	0.0015 &	SOPHIE+ \\
3884.4642 &	-15.2776 &	0.0042 &	+30.30 &	0.1095 &	0.0015 &	SOPHIE+ \\
3903.5775 &	-15.2705 &	0.0049 &	+19.70 &	0.1128 &	0.0013 &	SOPHIE+ \\
4150.5838 &	-15.3645 &	0.0042 &	+15.80 &	0.1108 &	0.0015 &	SOPHIE+ \\
4153.6009 &	-15.3735 &	0.0042 &	+24.70 &	0.1101 &	0.0015 &	SOPHIE+ \\
\end{longtable}

%% file: HD214823_rvtable.tex
\begin{longtable}{r r r r r r l}
\caption{Radial velocity measurements of HD214823. \label{table.rvHD214823}}\\
\hline\hline
\multicolumn{1}{c}{BJD} &	\multicolumn{1}{c}{RV} &	\multicolumn{1}{c}{$\sigma_\text{RV}$} &	\multicolumn{1}{c}{BIS} &	\multicolumn{1}{c}{H$\alpha$} &	\multicolumn{1}{c}{$\sigma_{\text{H}\alpha}$}&	\multicolumn{1}{l}{Instrument}\\
-2 453 000  &(\kms) & (\kms) & (\ms) & & \\
\hline
\noalign{\smallskip}
\endfirsthead
\caption{Continued.}\\
\hline
\multicolumn{1}{c}{BJD} &	\multicolumn{1}{c}{RV} &	\multicolumn{1}{c}{$\sigma_\text{RV}$} &	\multicolumn{1}{c}{BIS} &	\multicolumn{1}{c}{H$\alpha$} &	\multicolumn{1}{c}{$\sigma_{\text{H}\alpha}$}&	\multicolumn{1}{l}{Instrument}\\
-2 453 000  &(\kms) & (\kms) & (\ms) & (\kms) & & \\
\hline
\noalign{\smallskip}
\endhead
\hline
\endfoot
\hline
\endlastfoot
 586.5866 &	-44.4791 &	0.0273 &	 +0.00 &	0.1120 &	\multicolumn{1}{c}{---} &	ELODIE \\
 934.5516 &	-44.8000 &	0.0207 &	 +0.00 &	0.1120 &	\multicolumn{1}{c}{---} &	ELODIE \\
 935.5938 &	-44.7740 &	0.0169 &	 +0.00 &	0.1130 &	\multicolumn{1}{c}{---} &	ELODIE \\
 936.5818 &	-44.8330 &	0.0149 &	 +0.00 &	0.1120 &	\multicolumn{1}{c}{---} &	ELODIE \\
 960.5884 &	-44.8050 &	0.0262 &	 +0.00 &	0.1170 &	\multicolumn{1}{c}{---} &	ELODIE \\
1667.6133 &	-44.4108 &	0.0051 &	+11.70 &	0.1115 &	0.0016 &	SOPHIE \\
2065.5018 &	-44.2597 &	0.0050 &	+13.80 &	0.1083 &	0.0016 &	SOPHIE \\
2066.5028 &	-44.2559 &	0.0059 &	+29.70 &	0.1071 &	0.0018 &	SOPHIE \\
2078.5658 &	-44.2532 &	0.0056 &	+62.30 &	0.1064 &	0.0017 &	SOPHIE \\
2079.4718 &	-44.2396 &	0.0074 &	+37.20 &	0.1071 &	0.0022 &	SOPHIE \\
2079.4766 &	-44.2157 &	0.0097 &	+16.00 &	0.1068 &	0.0029 &	SOPHIE \\
2081.5076 &	-44.2476 &	0.0055 &	+10.30 &	0.1086 &	0.0017 &	SOPHIE \\
2402.5997 &	-44.4194 &	0.0048 &	+18.80 &	0.1086 &	0.0015 &	SOPHIE \\
2403.5624 &	-44.3842 &	0.0059 &	+18.30 &	0.1083 &	0.0018 &	SOPHIE \\
2466.4048 &	-44.4696 &	0.0058 &	+26.50 &	0.1123 &	0.0018 &	SOPHIE \\
2557.2851 &	-44.6110 &	0.0060 &	+12.80 &	0.1136 &	0.0018 &	SOPHIE \\
2583.2928 &	-44.6219 &	0.0063 &	+26.00 &	0.1120 &	0.0016 &	SOPHIE \\
2702.6084 &	-44.7645 &	0.0054 &	+23.50 &	0.1098 &	0.0015 &	SOPHIE \\
2776.6010 &	-44.7816 &	0.0059 &	+27.80 &	0.1119 &	0.0018 &	SOPHIE+ \\
2848.3610 &	-44.8006 &	0.0051 &	+16.50 &	0.1107 &	0.0015 &	SOPHIE+ \\
2969.2450 &	-44.7824 &	0.0064 &	+12.50 &	0.1069 &	0.0016 &	SOPHIE+ \\
3058.6237 &	-44.7454 &	0.0060 &	+18.00 &	0.1082 &	0.0017 &	SOPHIE+ \\
3136.5924 &	-44.6834 &	0.0067 &	 +9.50 &	0.1103 &	0.0020 &	SOPHIE+ \\
3314.2602 &	-44.5371 &	0.0058 &	+32.70 &	0.1083 &	0.0016 &	SOPHIE+ \\
3495.5916 &	-44.4194 &	0.0058 &	+31.70 &	0.1077 &	0.0017 &	SOPHIE+ \\
3610.3853 &	-44.3440 &	0.0057 &	+17.80 &	0.1070 &	0.0017 &	SOPHIE+ \\
3850.5699 &	-44.2546 &	0.0059 &	 +8.80 &	0.1060 &	0.0017 &	SOPHIE+ \\
3990.3415 &	-44.2506 &	0.0054 &	+23.70 &	0.1062 &	0.0015 &	SOPHIE+ \\
4203.5919 &	-44.3272 &	0.0052 &	+13.50 &	0.1053 &	0.0016 &	SOPHIE+ \\
\end{longtable}

%% file: HD221585_rvtable.tex
\begin{longtable}{r r r r r r l}
\caption{Radial velocity measurements of HD221585. \label{table.rvHD221585}}\\
\hline\hline
\multicolumn{1}{c}{BJD} &	\multicolumn{1}{c}{RV} &	\multicolumn{1}{c}{$\sigma_\text{RV}$} &	\multicolumn{1}{c}{BIS} &	\multicolumn{1}{c}{H$\alpha$} &	\multicolumn{1}{c}{$\sigma_{\text{H}\alpha}$}&	\multicolumn{1}{l}{Instrument}\\
-2 453 000  &(\kms) & (\kms) & (\ms) & & \\
\hline
\noalign{\smallskip}
\endfirsthead
\caption{Continued.}\\
\hline
\multicolumn{1}{c}{BJD} &	\multicolumn{1}{c}{RV} &	\multicolumn{1}{c}{$\sigma_\text{RV}$} &	\multicolumn{1}{c}{BIS} &	\multicolumn{1}{c}{H$\alpha$} &	\multicolumn{1}{c}{$\sigma_{\text{H}\alpha}$}&	\multicolumn{1}{l}{Instrument}\\
-2 453 000  &(\kms) & (\kms) & (\ms) & (\kms) & & \\
\hline
\noalign{\smallskip}
\endhead
\hline
\endfoot
\hline
\endlastfoot
 277.4772 &	6.3213 &	0.0078 &	 +0.00 &	0.1110 &	\multicolumn{1}{c}{---} &	ELODIE \\
 280.4344 &	6.3193 &	0.0077 &	 +0.00 &	0.1160 &	\multicolumn{1}{c}{---} &	ELODIE \\
 337.3191 &	6.3239 &	0.0078 &	 +0.00 &	0.1150 &	\multicolumn{1}{c}{---} &	ELODIE \\
 625.4868 &	6.3475 &	0.0090 &	 +0.00 &	0.1140 &	\multicolumn{1}{c}{---} &	ELODIE \\
 627.5011 &	6.3345 &	0.0076 &	 +0.00 &	0.1130 &	\multicolumn{1}{c}{---} &	ELODIE \\
 641.4412 &	6.3365 &	0.0078 &	 +0.00 &	0.1120 &	\multicolumn{1}{c}{---} &	ELODIE \\
 749.2843 &	6.3581 &	0.0108 &	 +0.00 &	0.1420 &	\multicolumn{1}{c}{---} &	ELODIE \\
 751.2521 &	6.3561 &	0.0131 &	 +0.00 &	0.1150 &	\multicolumn{1}{c}{---} &	ELODIE \\
 755.2365 &	6.3491 &	0.0081 &	 +0.00 &	0.1150 &	\multicolumn{1}{c}{---} &	ELODIE \\
 932.5780 &	6.3980 &	0.0103 &	 +0.00 &	0.1140 &	\multicolumn{1}{c}{---} &	ELODIE \\
1694.5785 &	6.4305 &	0.0033 &	-25.80 &	0.1185 &	0.0016 &	SOPHIE \\
1725.5496 &	6.4214 &	0.0032 &	-16.20 &	0.1130 &	0.0015 &	SOPHIE \\
1726.4688 &	6.4267 &	0.0029 &	-11.30 &	0.1174 &	0.0015 &	SOPHIE \\
1767.3362 &	6.4260 &	0.0032 &	 -4.00 &	0.1168 &	0.0015 &	SOPHIE \\
2429.5960 &	6.4153 &	0.0031 &	-16.80 &	0.1139 &	0.0015 &	SOPHIE \\
2790.6300 &	6.4103 &	0.0020 &	-13.50 &	0.1105 &	0.0009 &	SOPHIE+ \\
3115.6051 &	6.4547 &	0.0019 &	-12.70 &	0.1102 &	0.0008 &	SOPHIE+ \\
3174.5607 &	6.4651 &	0.0031 &	-21.00 &	0.1143 &	0.0015 &	SOPHIE+ \\
3314.2690 &	6.4501 &	0.0033 &	-22.70 &	0.1119 &	0.0015 &	SOPHIE+ \\
3329.2560 &	6.4607 &	0.0033 &	-17.80 &	0.1103 &	0.0015 &	SOPHIE+ \\
3466.5940 &	6.4407 &	0.0024 &	-21.70 &	0.1122 &	0.0011 &	SOPHIE+ \\
3518.6086 &	6.4323 &	0.0033 &	-14.70 &	0.1122 &	0.0016 &	SOPHIE+ \\
3519.5768 &	6.4310 &	0.0033 &	-25.80 &	0.1137 &	0.0016 &	SOPHIE+ \\
3520.5308 &	6.4348 &	0.0032 &	-19.50 &	0.1115 &	0.0016 &	SOPHIE+ \\
3587.3201 &	6.4221 &	0.0033 &	-29.20 &	0.1124 &	0.0015 &	SOPHIE+ \\
3612.2785 &	6.4208 &	0.0058 &	-15.00 &	0.1122 &	0.0027 &	SOPHIE+ \\
3612.2849 &	6.4230 &	0.0089 &	-33.00 &	0.1123 &	0.0040 &	SOPHIE+ \\
3612.3122 &	6.4166 &	0.0033 &	-28.00 &	0.1130 &	0.0016 &	SOPHIE+ \\
3847.5669 &	6.4095 &	0.0030 &	-22.20 &	0.1105 &	0.0015 &	SOPHIE+ \\
3941.2882 &	6.4056 &	0.0030 &	-21.20 &	0.1157 &	0.0014 &	SOPHIE+ \\
4026.2424 &	6.4126 &	0.0028 &	-31.70 &	0.1101 &	0.0014 &	SOPHIE+ \\
4053.2945 &	6.4214 &	0.0030 &	-25.70 &	0.1113 &	0.0013 &	SOPHIE+ \\
4184.6013 &	6.4421 &	0.0028 &	-14.20 &	0.1142 &	0.0012 &	SOPHIE+ \\
4202.5927 &	6.4359 &	0.0020 &	-21.80 &	0.1109 &	0.0009 &	SOPHIE+ \\
4259.4879 &	6.4463 &	0.0033 &	-22.70 &	0.1148 &	0.0016 &	SOPHIE+ \\
\end{longtable}

%% file: priors.tex
\begin{table*}[t]
{\small\centering
\caption{Parameter prior distributions. \label{table.priors}}            
\begin{tabular}{l l c c c c}        
\hline\hline                 
\noalign{\smallskip}

\multicolumn{2}{c}{ Orbital parameters } 	&\multicolumn{4}{c}{Prior distribution}\\
				&				&HD16175	&HD191806	&HD214823	&HD221585\\
\hline
\noalign{\smallskip}
Orbital period, $P$ 	&[days]	 		&\multicolumn{4}{c}{$\longleftarrow J(1, 10^4)\longrightarrow$}\\

RV amplitude, $K$ 	&[\ms]			&$U(0, 200)$	&$U(0, 200)$&$U(0, 400)$&$U(0, 200)$\\

Eccentricity, $e$    	& 				&\multicolumn{4}{c}{$\longleftarrow B(0.867, 3.03)\longrightarrow$}\\

Argument of periastron, $\omega$&[deg]	&\multicolumn{4}{c}{$\longleftarrow U(0, 360)\longrightarrow$}\\

Mean longitude at epoch, $L_0$ &[deg] 	&\multicolumn{4}{c}{$\longleftarrow U(0, 360)\longrightarrow$}\\
Systemic velocity, $V_0$		&[\kms]	&$U(21.75, 21.94)$	&$U(-18.65, -12.08)$ & $U(-44.56, -44.47)$ & $U(6.41, 6.46)$\\
Linear drift				&[\ms/yr]	&	--	&$U(-1,1)$	&--&--\\
\noalign{\smallskip}
\multicolumn{2}{c}{Noise model} 		\\
\hline
\noalign{\smallskip}
ELODIE additional noise			&[\ms]	&\multicolumn{4}{c}{$\longleftarrow U(0, 150)\longrightarrow$}\\	
SOPHIE noise at activity minimum   	&[\ms]	&	--	&$U(0, 50)$	&$U(0, 50)$	&$U(0, 50)$\\
SOPHIE+ noise at activity minimum 	&[\ms]	&\multicolumn{4}{c}{$\longleftarrow U(0, 50)\longrightarrow$}\\	
LICK additional noise			&[\ms]	&$U(0, 50)$	&--&--&--\\
SOPHIE/SOPHIE+ slope, $\alpha_J$&[\kms]	&\multicolumn{4}{c}{$\longleftarrow U(0, 10)\longrightarrow$}\\	
\noalign{\smallskip}
\multicolumn{2}{c}{Instrument offsets} 		\\
\hline
\noalign{\smallskip}

ELODIE/SOPHIE offset &[\ms]		&	--	&$N(166, 40)$	&$N(57.22, 23)$	&$N(112.5, 23)$\\
SOPHIE+/SOPHIE offset &[\ms]	&	--	&$N(0, 10)$	&$N(0, 10)$		&$N(0, 10)$\\
ELODIE/SOPHIE+ offset &[\ms]	&$N(69.98, 33)$	&--&--&--\\
LICK/SOPHIE+ offset &[\kms]		&$U(21.6,22.0)$	&--&--&--\\
\hline
\end{tabular}

\tablefoot{\\
Arrows indicate that the same prior applies to all stars.\\
$U(x_{min};  x_{max})$: uniform distribution between $x_{min}$ and $x_{max}$.\\
$J(x_{min};  x_{max})$: Jeffreys (log-flat) distribution between $x_{min}$ and $x_{max}$.\\
$N(\mu; \sigma)$: normal distribution with mean $\mu$ and standard deviation $\sigma$.\\
$B(a, b)$: beta distribution.\\
}

}
\end{table*}